\documentclass[twocolumn]{aastex631}

\DeclareRobustCommand{\ion}[2]{%
\relax\ifmmode
\ifx\testbx\f@series
{\mathbf{#1\,\mathsc{#2}}}\else
{\mathrm{#1\,\mathsc{#2}}}\fi
\else\textup{#1\,{\mdseries\textsc{#2}}}%
\fi}

\shorttitle{Non-Force Free Initialised Flare Energetics}
\shortauthors{Bate et al.}
\graphicspath{{./}{}}

\usepackage{gensymb}
\usepackage{amsmath}
\usepackage{physics}
\usepackage{subfigure}
\usepackage{graphicx}
\DeclareMathAlphabet\mathbfcal{OMS}{cmsy}{b}{n}

\begin{document}

\title{Energetics and Emission in a Simulated Solar Flare Initialised by a Non-Force Free Magnetic Field}

\correspondingauthor{W.~Bate}
\email{w.bate@herts.ac.uk}

\author[0000-0001-9629-5250]{W.~Bate}
\affiliation{Department of Physics, Astronomy \& Mathematics, University of Hertfordshire, Hatfield, AL10 9AB, UK}

\author[0000-0003-2291-4922]{M.~Gordovskyy}
\affiliation{Department of Physics, Astronomy \& Mathematics, University of Hertfordshire, Hatfield, AL10 9AB, UK}

\author[0000-0003-0819-464X]{A.~Prasad}
\affiliation{Statkraft AS, Lysaker, Norway}
\affiliation{Rosseland Centre for Solar Physics, University of Oslo, Blindern Postboks 1029, 0315 Oslo, Norway}
\affiliation{Institute of Theoretical Astrophysics, University of Oslo, Blindern Postboks 1029, 0315 Oslo, Norway}

\author[0000-0002-1729-8267]{A.~S.~Brun}
\affiliation{Universit\'{e} Paris-Saclay, Universit\'{e} Paris Cit\'{e}, CEA, CNRS, AIM, 91191 Gif-sur-Yvette, France}

\author[0000-0002-9630-6463]{A.~Strugarek}
\affiliation{Universit\'{e} Paris-Saclay, Universit\'{e} Paris Cit\'{e}, CEA, CNRS, AIM, 91191 Gif-sur-Yvette, France}

\author[0000-0002-1536-8508]{M.~V.~Sieyra}
\affiliation{Universit\'{e} Paris-Saclay, Universit\'{e} Paris Cit\'{e}, CEA, CNRS, AIM, 91191 Gif-sur-Yvette, France}

\author[0000-0002-7089-5562]{P.~K.~Browning}
\affiliation{Jodrell Bank Centre for Astrophysics, University of Manchester, Manchester, M13 9PL, UK}

\author[0000-0001-5121-5122]{S.~Inoue}
\affiliation{Center for Solar-Terrestrial Research, New Jersey Institute of Technology, University Heights, Newark, NJ 07102-1982, USA}

\author[0000-0003-2002-0247]{K.~Matsumoto}
\affiliation{Center for Solar-Terrestrial Research, New Jersey Institute of Technology, University Heights, Newark, NJ 07102-1982, USA}

\author[0000-0003-1363-3096]{A.~Roddanavar}
\affiliation{Center for Solar-Terrestrial Research, New Jersey Institute of Technology, University Heights, Newark, NJ 07102-1982, USA}



\begin{abstract}

Solar flare simulations are commonly initialised using non-linear force free field (NLFF) extrapolations derived from photospheric vector magnetograms. However, the force free assumption neglects plasma forces and may limit the available free magnetic energy. In this work, we perform a controlled comparison of two three-dimensional resistive magnetohydrodynamic simulations of the X2.1-class flare that occurred on 2011 September 06 in NOAA Active Region 11283. The simulations differ only in their initial magnetic configuration: one is based on a conventional NLFF extrapolation, while the other employs a non-force free extrapolation. Both models are evolved in an identical stratified atmosphere using the same numerical framework, enabling direct assessment of how the initial magnetic assumptions influence flare dynamics and energetics.

We find that the non-force free model undergoes more extensive magnetic restructuring and releases approximately twice as much magnetic energy ($\approx4.4 \times 10^{31}$~erg) as the NLFF case ($\approx2.3 \times 10^{31}$~erg), bringing the energy budget into closer agreement with expectations for X-class flares. Synthetic extreme ultraviolet emission in the 94~\AA~channel is computed for both simulations and compared with observations from the Solar Dynamics Observatory. The non-force free model produces a brighter and more spatially extended emission structure that more closely resembles the observed flare morphology and light curve. These results demonstrate that assumptions made in constructing the pre-flare coronal magnetic field can significantly affect flare energetics and observable signatures, and suggest that non-force free extrapolations provide a promising pathway toward more realistic data-constrained flare modelling.

\end{abstract}

\keywords{Solar atmosphere --- Solar corona --- Solar flares}


\section{Introduction} 
\label{sec:intro}

Solar flares are manifestations of rapid magnetic energy release in the solar corona. They are accompanied by strong plasma heating, bulk flows, and intense radiation across the electromagnetic spectrum. Extreme ultraviolet (EUV) and X-ray observations show that flaring coronal plasma is structured by the magnetic field and evolves rapidly during the energy release process, reflecting both the pre-flare magnetic configuration and its subsequent reconfiguration through magnetic reconnection. Understanding how the coronal magnetic field stores energy and how this energy is converted into thermal and nonthermal plasma during flares remain some of the central problems in solar physics.

Due to the fact that direct measurements of the coronal magnetic field are limited, numerical modelling plays a crucial role in interpreting flare observations \citep{2012LRSP....9....5W}. A common approach in flare modelling is to initialise magnetohydrodynamic (MHD) simulations using magnetic field extrapolations based on photospheric vector magnetograms. In particular, non-linear force free field (NLFF) extrapolations have been widely used in previous studies, motivated by the low plasma beta of the pre-flare corona and the assumption that Lorentz forces are small prior to flare onset \citep{gary_plasma_2001}. MHD models based on NLFF initial conditions have successfully reproduced a range of observed flare-related phenomena, including the large-scale evolution of coronal loops, current sheet formation, and locations of energy release \citep[see][for a comprehensive review]{toriumi_flare-productive_2019}.

However, the force free assumption is not strictly valid in the photosphere where the initial magnetograms are taken from. This is due to the fact that there is a non-negligible Lorentz force in the photosphere as plasma beta is on the order of unity \citep{gary_plasma_2001}. This requires significant preprocessing of the original photospheric magnetogram before performing the extrapolation, approximating the chromospheric magnetic field \citep{wiegelmann_can_2008,schrijver_nonlinear_2008}. This raises the question of how sensitive flare simulations and their observational signatures are to the choice of magnetic field model used to initialise the MHD calculations. 

A number of recent studies have demonstrated the power of data-constrained and data-driven MHD simulations in reproducing observed flare dynamics and energetics. For example, \citet{gordovskyy_forward_2020} performed three-dimensional MHD simulations of an X-class flare initialised with an NLFF extrapolation derived from Helioseismic and Magnetic Imager \citep[HMI;][]{2012SoPh..275..327S} vector magnetograms, showing that the subsequent evolution naturally produced the formation of current sheets, magnetic reconnection, and large-scale restructuring consistent with extreme ultraviolet (EUV) and X-ray observations. They highlighted, however, that NLFF-based models can struggle to supply sufficient free magnetic energy to fully account for the observed flare energy budget. Extending this approach, \citet{gordovskyy_particle_2023} incorporated more realistic thermodynamics and particle acceleration diagnostics, enabling direct comparison of synthetic emission and non-thermal signatures with observations, and further emphasising the sensitivity of flare evolution to the pre-eruptive magnetic configuration. Similarly, \citet{masson_flux_2017} modelled a confined flare using a magnetofrictional relaxation followed by full MHD evolution, successfully reproducing the development of quasi-separatrix layers, slipping reconnection, and flare ribbon morphology. Together, these works illustrate that while NLFF-initialised MHD models can capture many key aspects of flare topology and reconnection dynamics, the detailed energy release, emission structure, and reconnection geometry depend critically on the assumed pre-flare magnetic field, motivating continued investigation of alternative extrapolation strategies such as the non-force free approach adopted within this work.

In addition to the choice of magnetic field extrapolation, the thermodynamic structure of the background atmosphere plays a significant role in flare simulations. Early three-dimensional MHD studies of eruptive events often adopted uniform density and pressure profiles or zero-beta approximations to simplify the dynamics and isolate magnetic effects \citep[e.g.][]{1994ASPC...68..225M}. However, the solar corona is strongly stratified under gravity, with density and pressure decreasing exponentially with height above the photosphere \citep[e.g.][]{gary_plasma_2001}. More recent flare and flux-rope simulations have therefore incorporated gravitationally stratified atmospheres to better capture the variation of Alfv\'{e}n speed, plasma beta, and energy transport with height \citep[e.g.][]{gordovskyy_particle_2014}. A stratified atmosphere provides a more realistic distribution of mass loading along magnetic field lines, influences the onset and growth of instabilities, and affects the partition of released magnetic energy into kinetic and thermal components \citep{Sieyra2026}. It also enables more meaningful forward modelling of optically thin EUV and X-ray emission, since the emissivity depends sensitively on the local density squared. Consequently, including stratification is important not only for reproducing the large-scale magnetic evolution, but also for obtaining quantitatively reliable synthetic observables and flare energetics.

Alternative approaches have been developed that relax the force free constraint and allow for magnetic field configurations that are not in force balance, incorporating plasma forces self-consistently from the outset \citep[][]{hu_improved_2008,hu_practical_2008,hu_non-force-free_2010}. These non-force free models aim to provide a more realistic representation of the low-coronal magnetic field and plasma state, potentially leading to different flare dynamics and observable signatures. Recently, these non-force free extrapolations have been used to initialise simulations leading to solar flares, coronal jets, and coronal dimmings \citep[][]{prasad_magnetohydrodynamic_2018,prasad_magnetohydrodynamic_2020,nayak_data-constrained_2019, kumar_magnetohydrodynamics_2022,prasad_formation_flux_rope_2023, Sieyra2026}. They also offer an opportunity to assess the limitations of traditional force free extrapolations and to evaluate whether a more complete treatment of the pre-flare coronal environment improves agreement with observations.

In this study, we directly compare two three-dimensional MHD models of the same solar flare that differ only in their initial magnetic field configuration. The first model is initialised using a standard NLFF extrapolation, representative of the approach commonly adopted in previous flare modelling studies. The second model employs a newly developed non-force free magnetic field model, which allows for finite Lorentz forces and includes plasma pressure and gravity in the initial equilibrium. Both models are evolved using the same MHD framework, enabling a controlled comparison of their magnetic evolution, plasma heating, and resulting coronal emission.

To evaluate the realism of the two modelling approaches, we compute synthetic extreme ultraviolet emission in the Atmospheric Imaging Assembly \citep[AIA;][]{2012SoPh..275...17L} $94$~\AA~channel from each simulation and compare it with observations of the actual flare. The $94$~\AA~channel is particularly sensitive to hot flare plasma and provides a stringent test of the models’ ability to reproduce the spatial structure and temporal evolution of flare-heated coronal loops. By comparing synthetic and observed EUV emission, we assess how differences in the initial magnetic field assumptions propagate into observable consequences during the flare.

The paper is organised as follows. 

Section~\ref{sec:obs} details the observations and extrapolations used within this work, including the NLFF extrapolation (\ref{sub:NLFF}); non-force free extrapolation (\ref{sub:jxb}); and the production of synthetic AIA observations (\ref{sub:synth}). Section~\ref{sec:sim} details the simulation setup and comparisons between the two initial fields; Section~\ref{sec:res} presents the results and analysis; and Section~\ref{sec:conc} details the authors' conclusions.

\section{Observations and Extrapolations}
\label{sec:obs}

For this work, observations of the X2.1 class flare on 2011 September 06 in Active Region AR11283 were utilised. These observations were taken with the the Solar Dynamics Observatory's \citep[SDO;][]{2012SoPh..275....3P} Helioseismic and Magnetic Imager \citep[HMI;][]{2012SoPh..275..327S} and Atmospheric Imaging Assembly \citep[AIA;][]{2012SoPh..275...17L}. This flare reached peak brightness at 22:21~UT and a pre-flare HMI magnetogram was used, taken at 20:48~UT. A section of this HMI magnetogram was taken using the Space-weather HMI Active Region Patch \citep[SHARP;][]{2014SoPh..289.3549B} for SHARP $833$ (associated with AR11283). The cylindrical equal area (CEA) projected version of this vector magnetogram was used to perform the magnetic field extrapolations described below. Further, AIA observations from the $94$~\AA~channel are taken due to the intensity of this flare and the fact that the peak of the $94$~\AA~response function is at the highest temperature of any of the AIA channels \citep[Table 1 of][]{2012SoPh..275...17L}.

\subsection{Non-linear Force Free Extrapolation}
\label{sub:NLFF}

The NLFF extrapolation was obtained using the methods described in \citet{inoue_nonlinear_2013}. This method is based on magnetic relaxation, developed by \citet{1994ApJ...422..899M,1994ASPC...68..225M}, with an additional algorithm to prevent deviation from the $\nabla\cdot\mathbf{B}=0$ condition. Firstly, the zero-beta (without gas pressure and gravity) MHD equations are considered. Therefore, the following equations are numerically solved,

\begin{equation}
    \frac{\partial \mathbf{v}}{\partial t} = -(\mathbf{v}\cdot\nabla)\mathbf{v} + \frac{1}{\rho} (\mathbf{J}\times\mathbf{B}) + \nu\nabla^{2}\mathbf{v} \ ,
\end{equation}
\begin{equation}
\label{eqn:induction}
    \frac{\partial\mathbf{B}}{\partial t} = \nabla\times((\mathbf{v}\times\mathbf{B})-\eta\mathbf{J})-\nabla\phi \ ,
\end{equation}
\begin{equation}
    \mathbf{J}=\nabla\times\mathbf{B} \ ,
\end{equation}

where $\mathbf{B}$ is the magnetic flux density, $\mathbf{v}$ is the velocity, $\mathbf{J}$ is the electric current density, $\rho$ is the pseudo density, $\phi$ is the initial potential field, $\nu$ is the viscosity, and $\eta$ is the resistivity. The pseudo density is set as proportional to $|\mathbf{B}|$ in order to make the Alfv\'{e}n speed equal everywhere, easing the relaxation process.

The additional step to prevent deviations from $\nabla\cdot\mathbf{B}=0$ was first introduced by \citet{dedner_hyperbolic_2002} and is as follows,

\begin{equation}
\label{eqn:dedner}
    \frac{\partial\phi}{\partial t} + c_{h}^{2} (\nabla\cdot\mathbf{B}) = -\frac{c_{h}^{2}}{c_{p}^{2}} \phi \ ,
\end{equation}

where the combination of Equations~\ref{eqn:induction} and \ref{eqn:dedner} leads to, 

\begin{equation}
    \frac{\partial^{2}(\nabla\cdot\mathbf{B})}{\partial t^{2}} + \frac{c_{h}^{2}}{c_{p}^{2}} \frac{\partial(\nabla\cdot\mathbf{B})}{\partial t} = c_{h}^{2} \nabla^{2}(\nabla\cdot\mathbf{B}) \ , 
\end{equation}

illustrating the propagating and diffusing nature of numerical errors relating to $\nabla\cdot\mathbf{B}$, where $c_{h}$ and $c_{p}$ correspond to the advection and diffusion coefficients, respectively. Following these equations and boundary conditions based on the observed vector magnetogram at the lower boundary, and rigid wall conditions for the others, the magnetic field is allowed to relax into a force free state.

\begin{figure*}[t]

    \centering

    \begin{subfigure}{}
        \centering
        \includegraphics[width=0.45\textwidth]{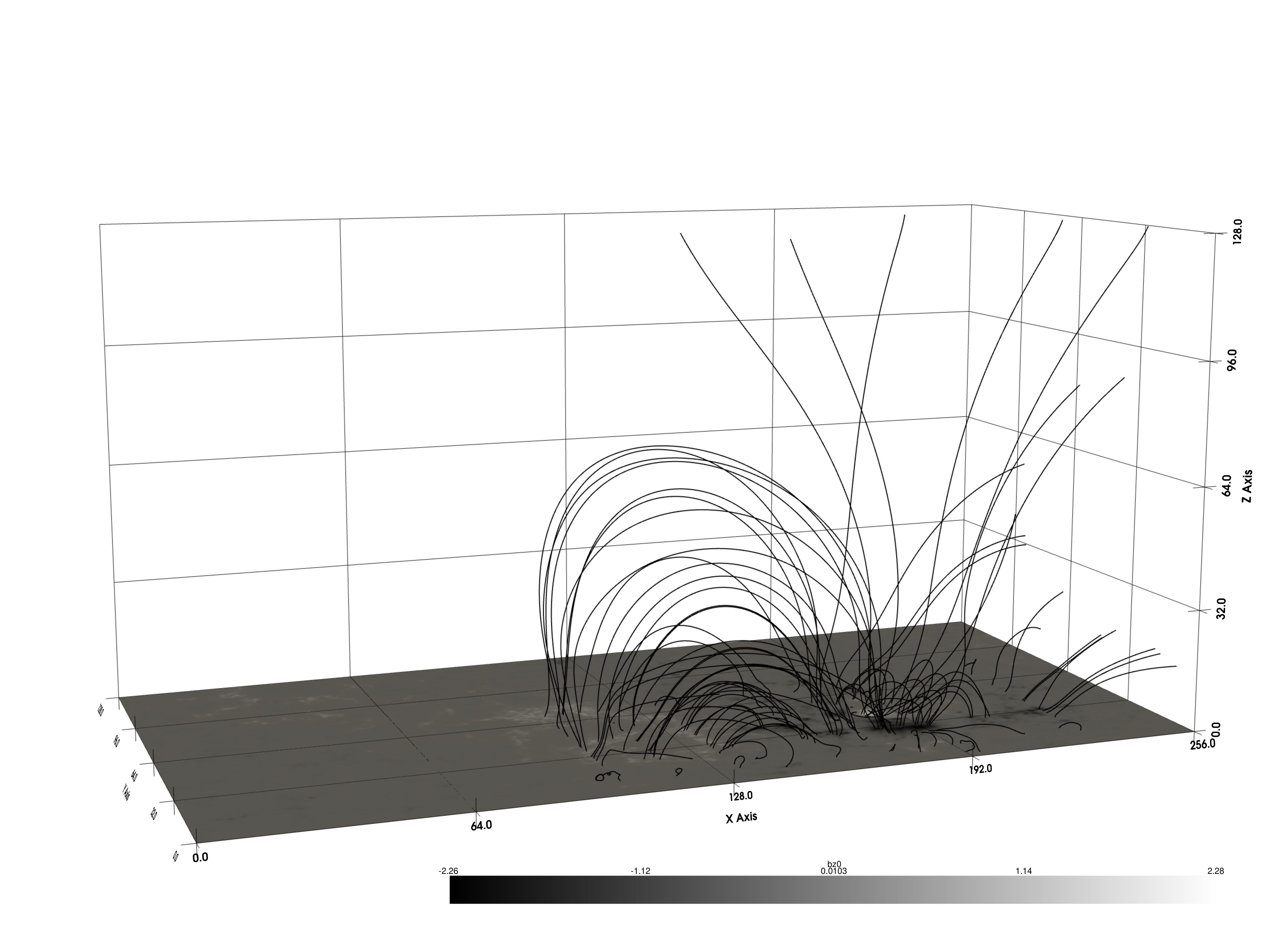}
    \end{subfigure}%
    \begin{subfigure}{}
        \centering
        \includegraphics[width=0.45\textwidth]{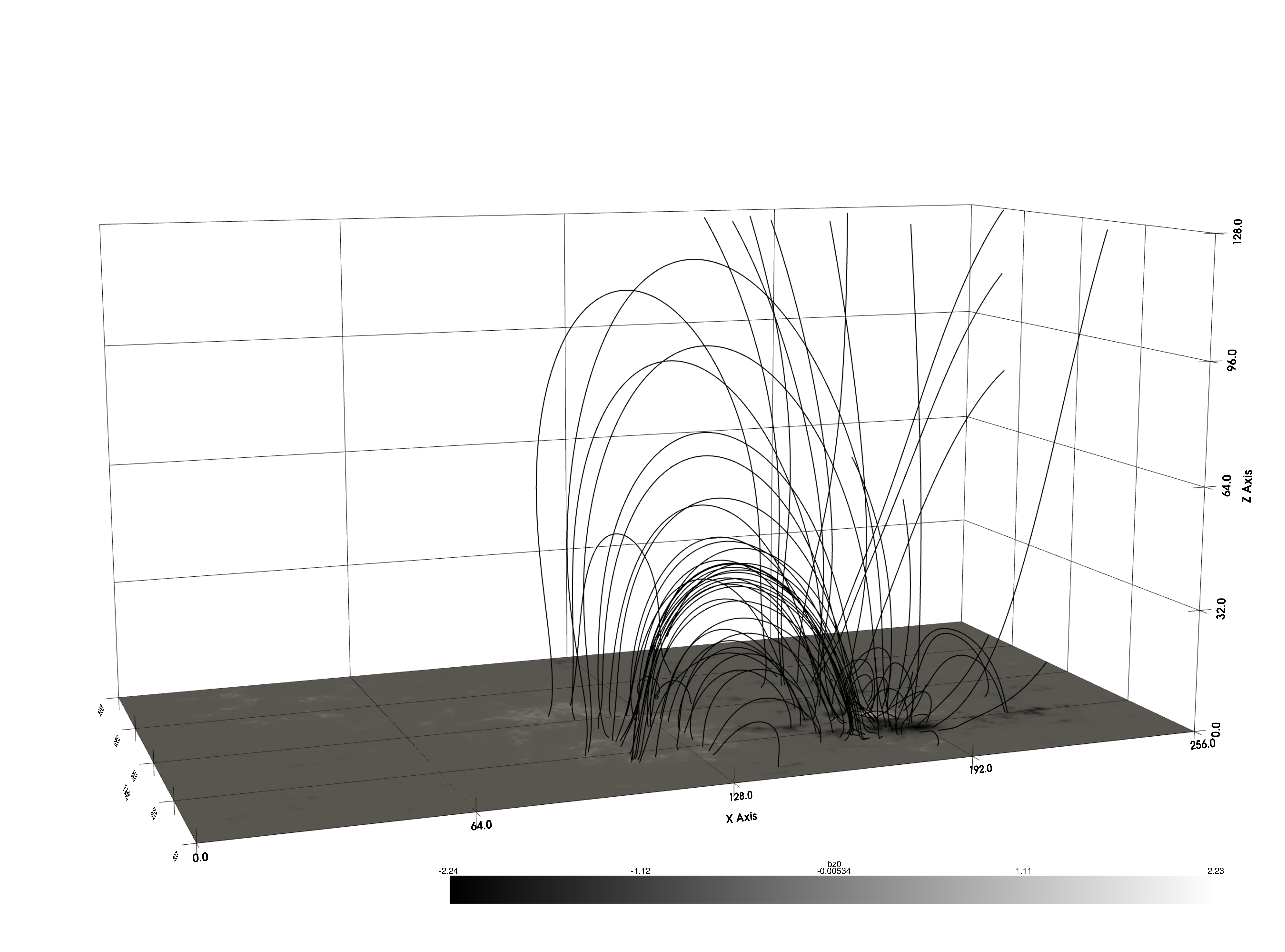}
    \end{subfigure}

    \begin{subfigure}{}
        \centering
        \includegraphics[width=0.45\textwidth]{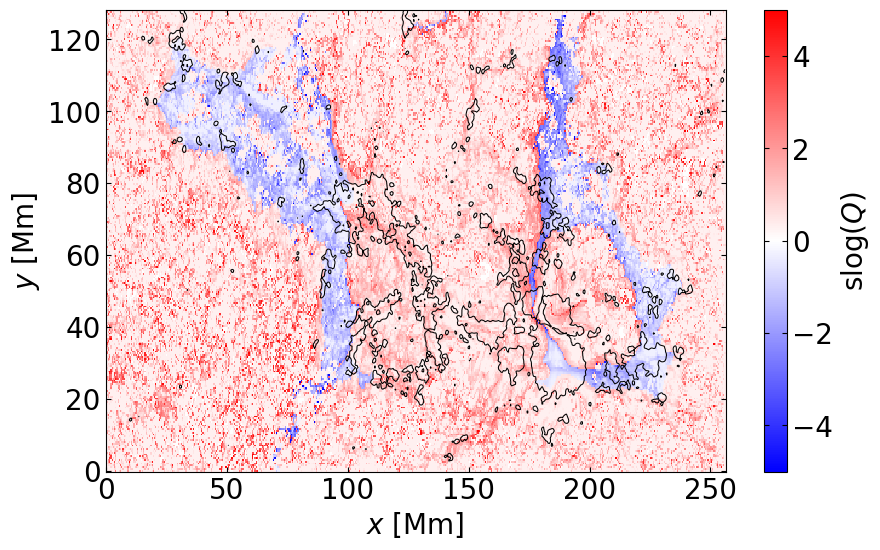}
    \end{subfigure}
    \begin{subfigure}{}
        \centering
        \includegraphics[width=0.45\textwidth]{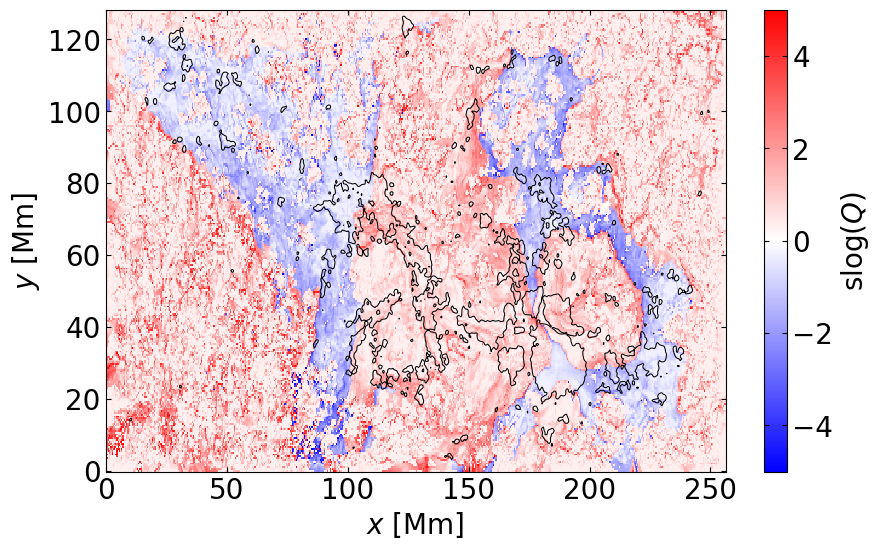}
    \end{subfigure}
    
    \begin{subfigure}{}
        \centering
        \includegraphics[width=0.45\textwidth]{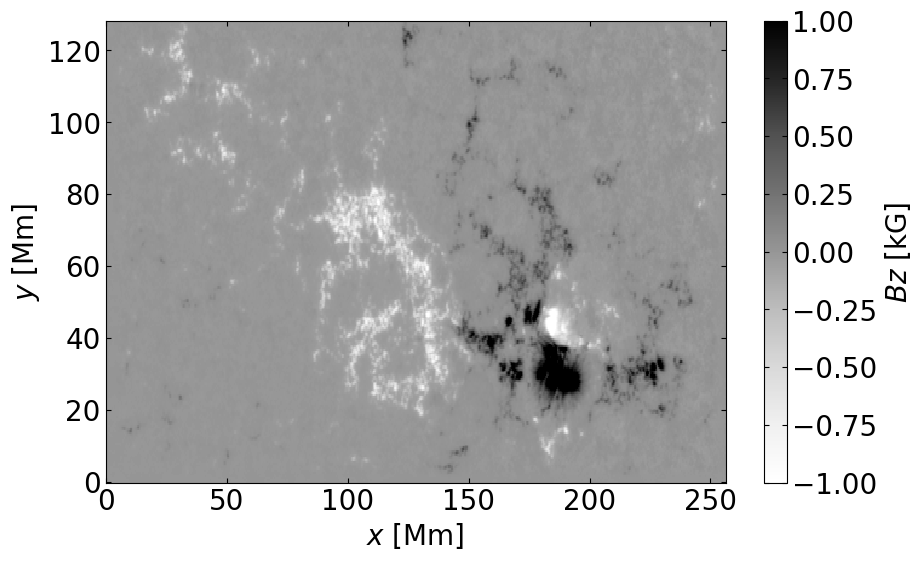}
    \end{subfigure}
    
    \caption{In the left column, the NLFF extrapolation is shown and in the right, the non-force free. The magnetic field lines shown are seeded at the same locations in both, with the field line density proportional to magnetic field strength at the lower boundary. The lower surface in the top row is coloured according to $B_{z}$ at the lower boundary. The second row shows the value of slogQ calculated at the lower boundary for the NLFF and non-force free field extrapolations respectively. Blue corresponds to open field lines, and red to closed lines. The black contour shows an absolute value of vertical magnetic field of $0.2$~kG. The lower panel shows the vertical magnetic field at the lower boundary, the initial surface magnetogram of AR11283, taken on 2011 September 06 at 20:48~UTC from SDO/HMI. This has been saturated at $\pm1$~kG for ease of comparison.}
    \label{fig:HMI}
\end{figure*}

\subsection{Non-force Free Extrapolation}
\label{sub:jxb}

The non-force free extrapolation was calculated using the method developed by \citet{hu_improved_2008,hu_practical_2008,hu_non-force-free_2010}. This approach combines three linear force free fields as

\begin{equation}
    \mathbf{B}=\mathbf{B}_{1}+\mathbf{B}_{2}+\mathbf{B}_{3} \ ;~\nabla\times\mathbf{B}_{i}=\alpha_{i}\mathbf{B}_{i} \ ,
\end{equation}

with $i=1,2,3$ and $\alpha_{i}$ the corresponding constant for each linear force free field.

$\mathbf{B}_2$ is chosen as a potential field, so $\alpha_2=0$, and we set $\alpha_1\neq\alpha_3$. An optimal set of $\alpha_1$ and $\alpha_3$ is found by minimising the difference between the observed and calculated transverse fields on the photospheric boundary ($\mathbf{B}_{t}$ and $\mathbf{b}_{t}$ respectively). The metric used to assess this is described in \citet{prasad_magnetohydrodynamic_2018} and is as follows, 

\begin{equation}
\label{eqn:E_n}
    E_{n}=\bigg(\sum_{i=1}^{M} |\mathbf{B}_{t,i}-\mathbf{b}_{t,i}| \times|\mathbf{B}_{t,i}|\bigg)\bigg/\bigg(\sum_{i=1}^{M} |\mathbf{B}_{t,i}|^{2}\bigg) \ ,
\end{equation}

where $M$ is the number of points in the photospheric plane. The denominator, normalising by the observed transverse magnetic field, ensures that contribution from weak fields is minimised \citep{hu_improved_2008,hu_non-force-free_2010}. 

This extrapolated field, $\mathbf{B}$, is the solution of the following equation

\begin{equation}
\label{eqn:curl}
    \nabla\times\nabla\times\nabla\times\mathbf{B} + a_{1}\nabla\times\nabla\times\mathbf{B} + b_{1}\nabla\times\mathbf{B}=0 \ ,
\end{equation}

which is a higher-curl equation with $a_{1},b_{1}$ as constants. This is derived by applying the principle of minimum dissipation rate (MDR), the process for which is described by \citet{bhattacharyya_dissipative_2004}, and Equation~\ref{eqn:curl} results from taking the curl of both sides of their Equation~12. Equation~\ref{eqn:curl} contains a second-order derivative at $z=0$, which would require magnetograms at two different $z$ values. In reality, only one photospheric magnetogram at a single height is typically available due to the use of e.g. HMI. In order to circumvent this, an iterative approach, described fully by \citet{hu_non-force-free_2010}, is used to minimise the deviation between the observed and the calculated transverse magnetic field (measured by $E_{n}$ in Equation~\ref{eqn:E_n}).

\begin{figure}
    \centering
    \begin{subfigure}{}
        \centering
        \includegraphics[width=0.95\linewidth]{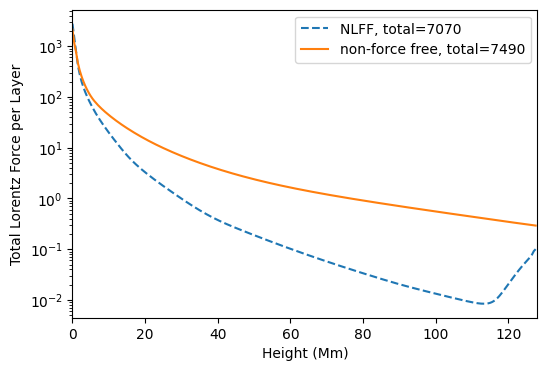}
    \end{subfigure}
    \begin{subfigure}{}
        \centering
        \includegraphics[width=0.95\linewidth]{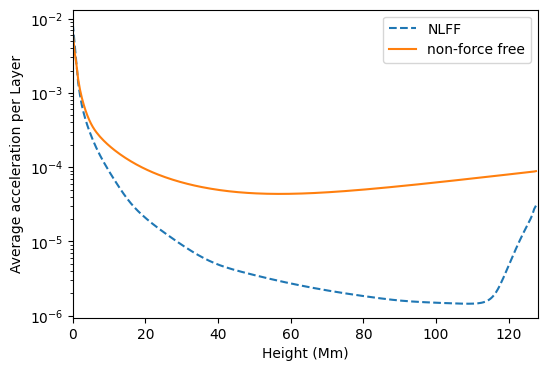}
    \end{subfigure}

    \caption{In the first panel, the Lorentz force summed in each layer of the extrapolations is shown on a log scale as a function of height. The solid line represents the NLFF model and the dashed line represents the non-force free model. In the second panel, this is recast as average acceleration per layer using force and density. Both dependent variables are presented in normalised units.} 
    \label{fig:lorentz}
\end{figure}

\subsection{Synthetic AIA Observations}
\label{sub:synth}

The Atmospheric Imaging Assembly \citep[AIA;][]{2012SoPh..275...17L} on board the Solar Dynamics Observatory \citep[SDO;][]{2012SoPh..275....3P} has proved to be an invaluable tool for modern solar physics studies since its launch in 2010. This is in part thanks to its $1.5''$ spatial resolution and $12$~s cadence with data taken over more than a decade. Further, the range of extreme ultraviolet (EUV) wavelengths available allows for the examination of a wide range of plasma temperatures. The wavelength with the highest characteristic $\log(T)$, with a value of $6.8$, is the $94$~\AA~channel, primarily focussed on the \ion{Fe}{xviii} ion. This is described as sampling the ``flaring corona'' and is thus particularly useful for flare studies \citep{2012SoPh..275...17L}.

In order to compare these AIA~$94$~\AA~observations to simulation results, synthetic observables are desired. The solar corona is optically thin to almost all forms of radiation, easing the creation of these synthetic observables \citep{morton_solar_2024}. First, the AIA response function for the $94$~\AA~channel is taken from Solar SoftWare \citep[SSW;][]{1998SoPh..182..497F}, calculated using CHIANTIv10 \citep{dere_chianti_1997,del_zanna_chiantiatomic_2021}, henceforth denoted as $\mathcal{R}$. The number density, $n$, and the temperature, $T$, are then taken directly from a snapshot of the MHD simulation. This allows for the calculation of the emissivity of each voxel as

\begin{equation}
\label{eqn:emissivity}
    \varepsilon(x,y,z)=n(x,y,z)^{2} \cdot \mathcal{R}\big(T(x,y,z)\big) \ .
\end{equation}

The optically thin nature of the solar corona then allows for simple integration of this emissivity along the line of sight to produce synthetic emission maps. 

\section{Simulations}
\label{sec:sim}

In this work, two simulations are produced for the Active Region AR11283 on 2011 September 06, which produced an X2.1 class flare, one utilising a NLFF extrapolation following \citet[][]{2014ApJ...780..101I} and described in Section~\ref{sub:NLFF}, and a second employing a non-force free extrapolation, following the approach developed by \citet{hu_improved_2008,hu_practical_2008,hu_non-force-free_2010,2009SoPh..257..271G} and described in Section~\ref{sub:jxb}. These extrapolations have a grid size of $512\times256\times256$, corresponding to a volume of $256\times128\times128$~Mm. The initial photospheric magnetogram from the Heliospheric Magnetic Imager \citep[HMI; ][]{2012SoPh..275..327S} is shown in Figure~\ref{fig:HMI}.


Each simulation is performed using the Lare3d code \citep{arber_staggered_2001} with the same grid as the extrapolations. The initial velocity is set as zero. The velocities at the lower boundary are held as zero throughout each simulation. The normalisation constants for length and magnetic flux density are $L_{0}=1.0$~Mm and $B_{0}=0.1$~T respectively. The normalisation constant for density, $\rho_{0}$, is initially held as a free parameter, which has no bearing on the results of the simulation, as all internal calculations are performed in normalised units. The value of $\rho_{0}$ only affects the interpretation of results when converting into physical units after the simulation has been run. For further discussion on the selection of the value of $\rho_{0}$, see Section~\ref{sec:res}. A snapshot is taken for each nominal Alfv\'{e}n time.

The Lare3d code simulates the plasma evolution according to the resistive MHD equations, presented here in normalised, Lagrangian forms, 

\begin{equation}
\label{eqn:momentum}
\frac{D\rho}{Dt} = -\rho  (\nabla\cdot\mathbf{v}) \ ,
\end{equation}

\begin{equation}
\label{}
\frac{\text{D}\mathbf{v}}{\text{D}t} = \frac{1}{\rho}(\nabla\times\mathbf{B})\times\mathbf{B}-\frac{1}{\rho}\nabla P - g\hat{\mathbf{z}} \ ,
\end{equation}

\begin{equation}
\label{}
\frac{D\mathbf{B}}{Dt} = (\mathbf{B}\cdot\nabla)\mathbf{v}-\mathbf{B}(\nabla\cdot\mathbf{v})-\nabla\times(\eta\nabla\times\mathbf{B}) \ ,
\end{equation}

\begin{equation}
\label{}
\frac{\text{D}\mathbf{\epsilon}}{\text{D}t} = -\frac{P}{\rho}\nabla\cdot\mathbf{v}+\frac{\eta}{\rho}J^{2} \ ,
\end{equation}

where $P$ is the gas pressure, $g$ is acceleration due to gravity, $\hat{\mathbf{z}}$ is the unit vector in the $z$-direction, and $\epsilon$ is the specific internal energy density.

A plane-parallel stratified atmosphere model is implemented at the beginning of each simulation. The density and pressure are stratified according to 
\begin{equation}
\label{eqn:density_prof}
    \rho=\rho_{a} e^{\frac{-z}{\Lambda}} \ , \
    P=P_{a} e^{\frac{-z}{\Lambda}} \ ,
\end{equation}

where $\rho_{a}$ is a constant, set at $2.5$; $P_{a}$ is a constant, set at $1/60$; and $\Lambda$ is the scale height, set as $27.8$ ($27.8$~Mm).

The specific internal energy density is then set at a constant value given by 
\begin{equation}
    \epsilon=\frac{P}{\rho(\gamma-1)} \ ,
\end{equation}
where $\gamma$ is the ratio of specific heats, set as $5/3$. This corresponds to an isothermal atmosphere as specific internal energy density is directly proportional to temperature. Initially, this results in a value of $\epsilon=1/100$ in normalised units. The value of $\rho_{0}$ chosen for production of the synthetic images corresponds to an initial temperature of $\approx0.55$~MK.

The gravity is set to act in the negative $z$-direction, and is set to balance out the pressure gradient term in the momentum equation (\ref{eqn:momentum}) such that,
\begin{equation}
    g=-\frac{1}{\rho}\nabla P=-\frac{1}{\rho_{a}}\frac{-1}{\Lambda}P_{a}\approx2.39\times10^{-4} \ .
\end{equation}

The resistivity, $\eta$, depends on the local plasma conditions is defined as follows:

\begin{equation}
\begin{aligned}
\eta&=\eta_{0}+\eta_{1} \ , \\  
\eta_{0}&=10^{-5} \ , \\
\eta_{1}&=
    \begin{cases}
      0 \ , & \text{if}\ j<\alpha\frac{\rho}{\sqrt{\epsilon}}~\mathrm{OR}~z<4  \\
      10^{-3} \ , & \text{if}\ j \geq \alpha\frac{\rho}{\sqrt{\epsilon}}~\mathrm{AND}~z\geq 4
    \end{cases}
    \ ,
\end{aligned}
\end{equation}
where $\alpha$ is a constant set as $0.002$, $\rho$ is the local density, and $\epsilon$ is the local specific internal energy density. This mimics the anomalous resistivity triggered by the ion-cyclotron instability \citep[e.g.][]{barta_spontaneous_2011,gordovskyy_particle_2014}.

The initial extrapolations for the NLFF and non-force free models are shown in the left and right panels of Figure~\ref{fig:HMI} respectively. In the first row, magnetic field lines seeded at the same locations for both extrapolations are plotted. In the second row, the signed logarithm of the squashing factor, $Q$, is shown at the lower boundary. The squashing factor is a measure of the deformation of the magnetic field line which passes through a particular point, formalised by \citet{titov_theory_2002}. These squashing factor values have been calculated using the The Universal Fieldline Tracer \citep[\texttt{UFiT};][]{aslanyan_new_2024} code. A positive value of squashing factor (shown in red) corresponds to a closed magnetic field line, and a negative value (shown in blue) to an open magnetic field line. This allows for the identification of different structures in the initial extrapolations, for example, both produce similar separatrix structures centred around $[x,y]=[200,50]$~Mm, characterised by a circle of red (closed field lines) surrounded by a ring of blue (open field lines). While $Q$ has a value everywhere, it is important to note that it is a purely geometric measure. A high $Q$ value does not necessarily imply a high field strength and vice-versa. To ease the interpretation of the $Q$ maps, the magnetic field strength at the lower boundary is shown in the bottom panel of Figure~\ref{fig:HMI}.

\begin{figure*}[th!]
\centering
\begin{subfigure}{}
    \centering
    \includegraphics[width=0.4\textwidth]{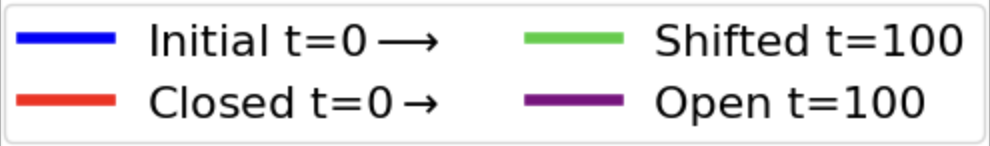}
\end{subfigure}%

\begin{subfigure}{}
    \centering
    \includegraphics[width=0.7\textwidth,trim={4cm 4.5cm 13cm 3cm},clip]{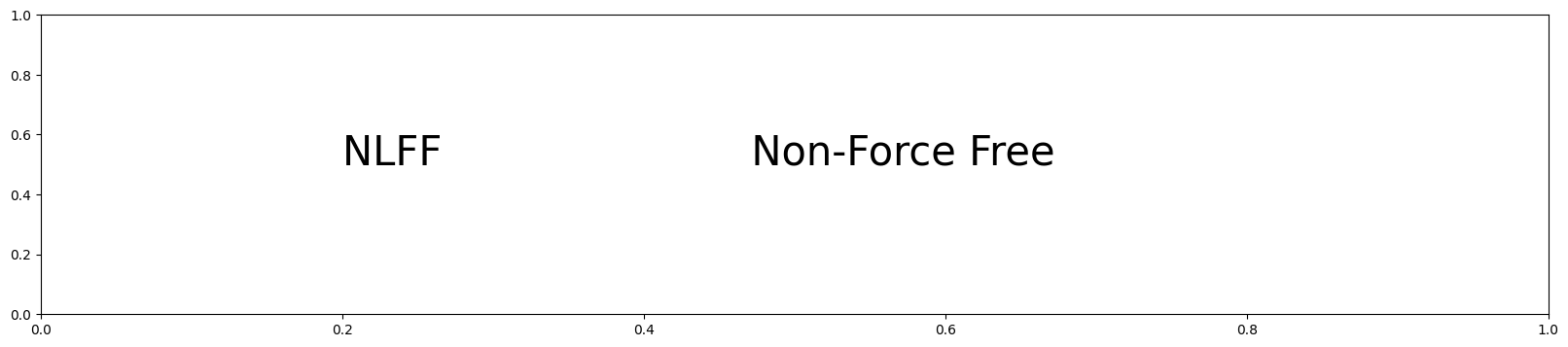}
\end{subfigure}%

\begin{subfigure}{}
    \centering
    \includegraphics[width=0.05\textwidth,trim={1.5cm 1cm 6.5cm 1.25cm},clip]{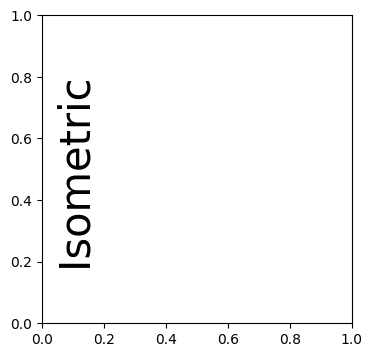}
\end{subfigure}%
\begin{subfigure}{}
    \centering
    \includegraphics[width=0.375\textwidth,trim={0cm 0cm 0cm 4cm},clip]{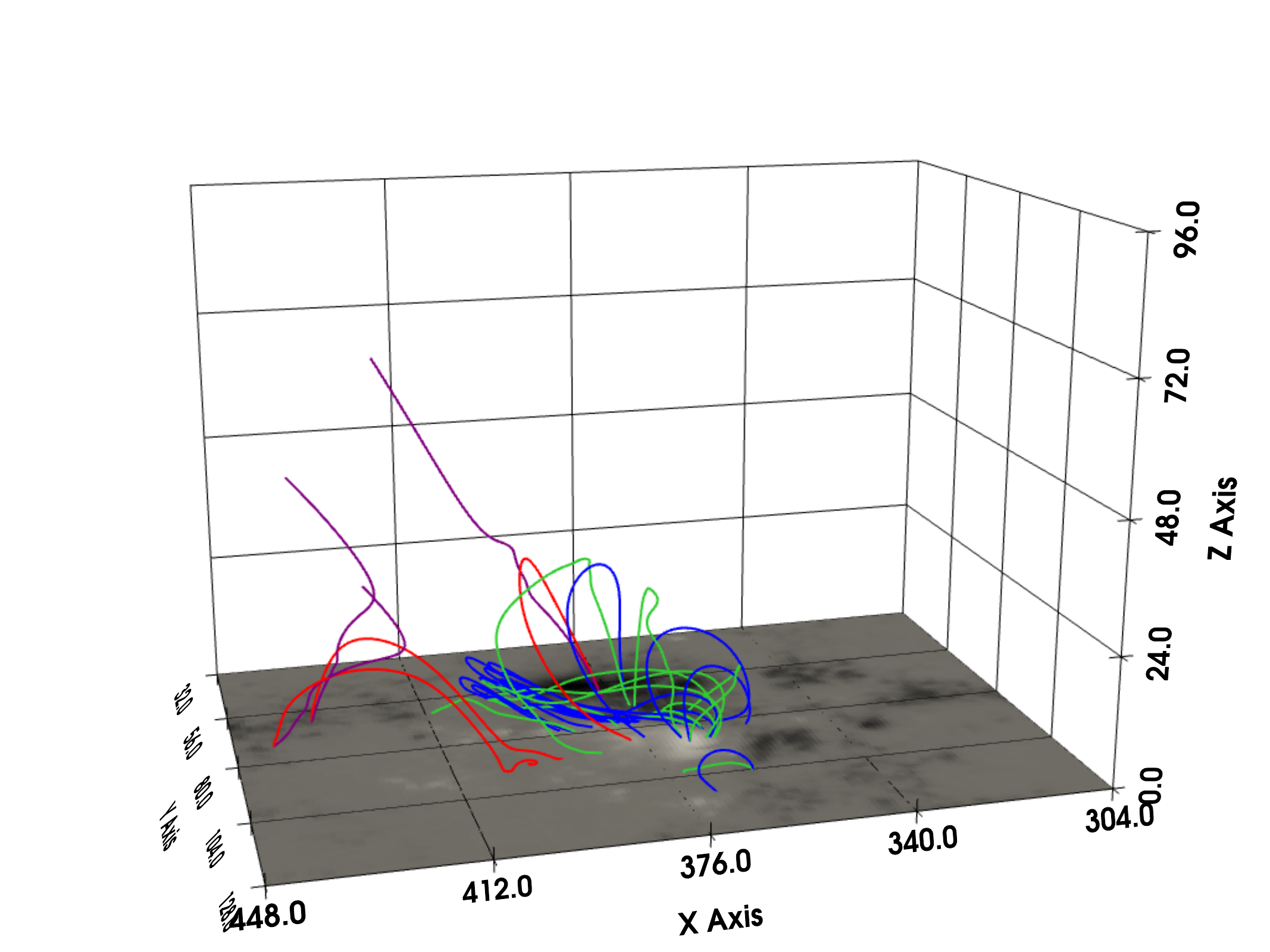}
\end{subfigure}%
\begin{subfigure}{}
    \centering
    \includegraphics[width=0.375\textwidth,trim={0cm 0cm 0cm 4cm},clip]{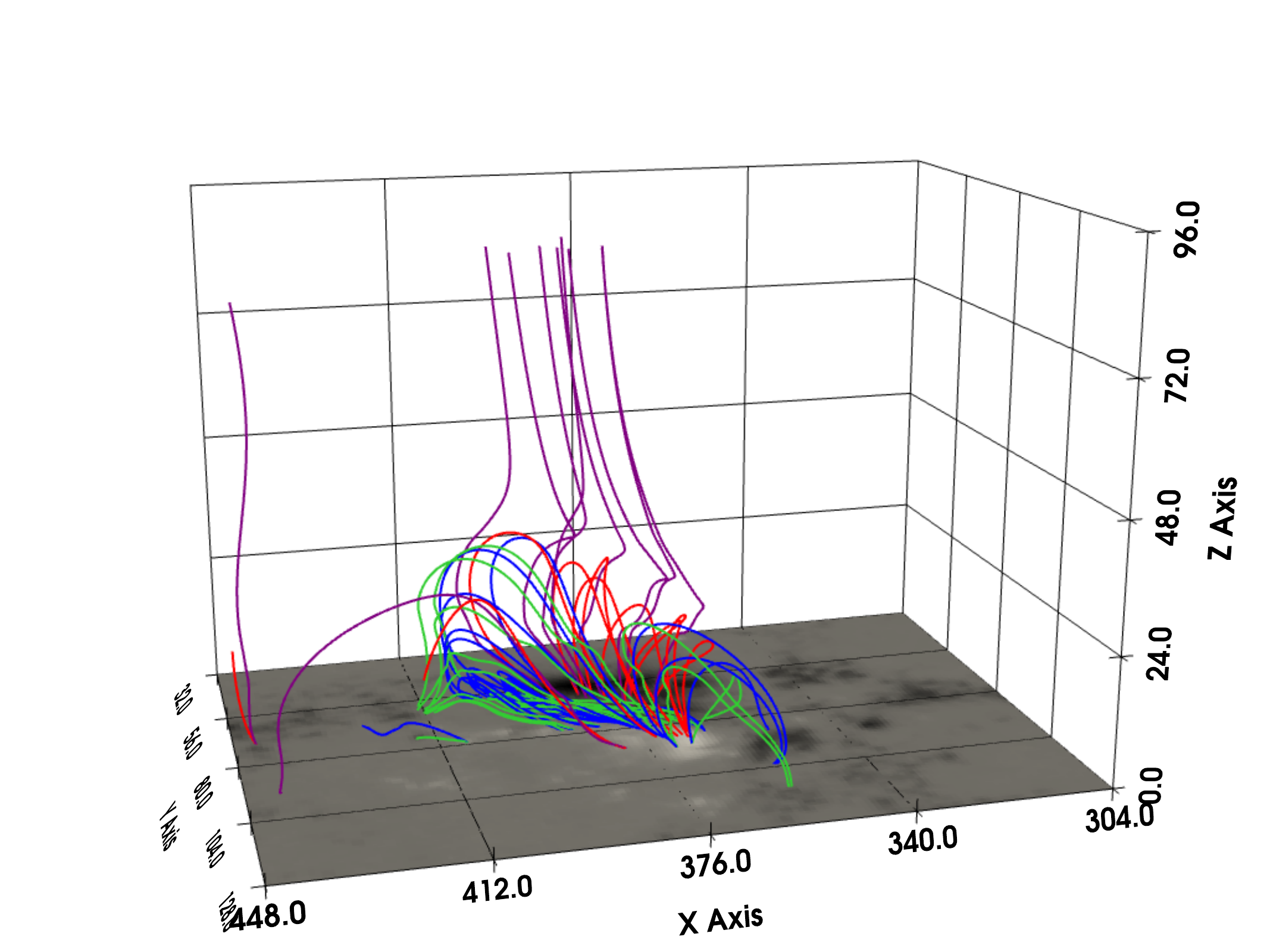}
\end{subfigure}

\begin{subfigure}{}
    \centering
    \includegraphics[width=0.05\textwidth,trim={1.5cm 1cm 6.5cm 1.25cm},clip]{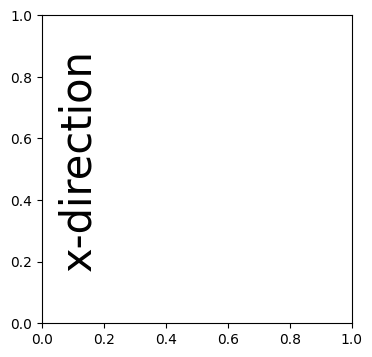}
\end{subfigure}%
\begin{subfigure}{}
    \centering
    \includegraphics[width=0.375\textwidth]{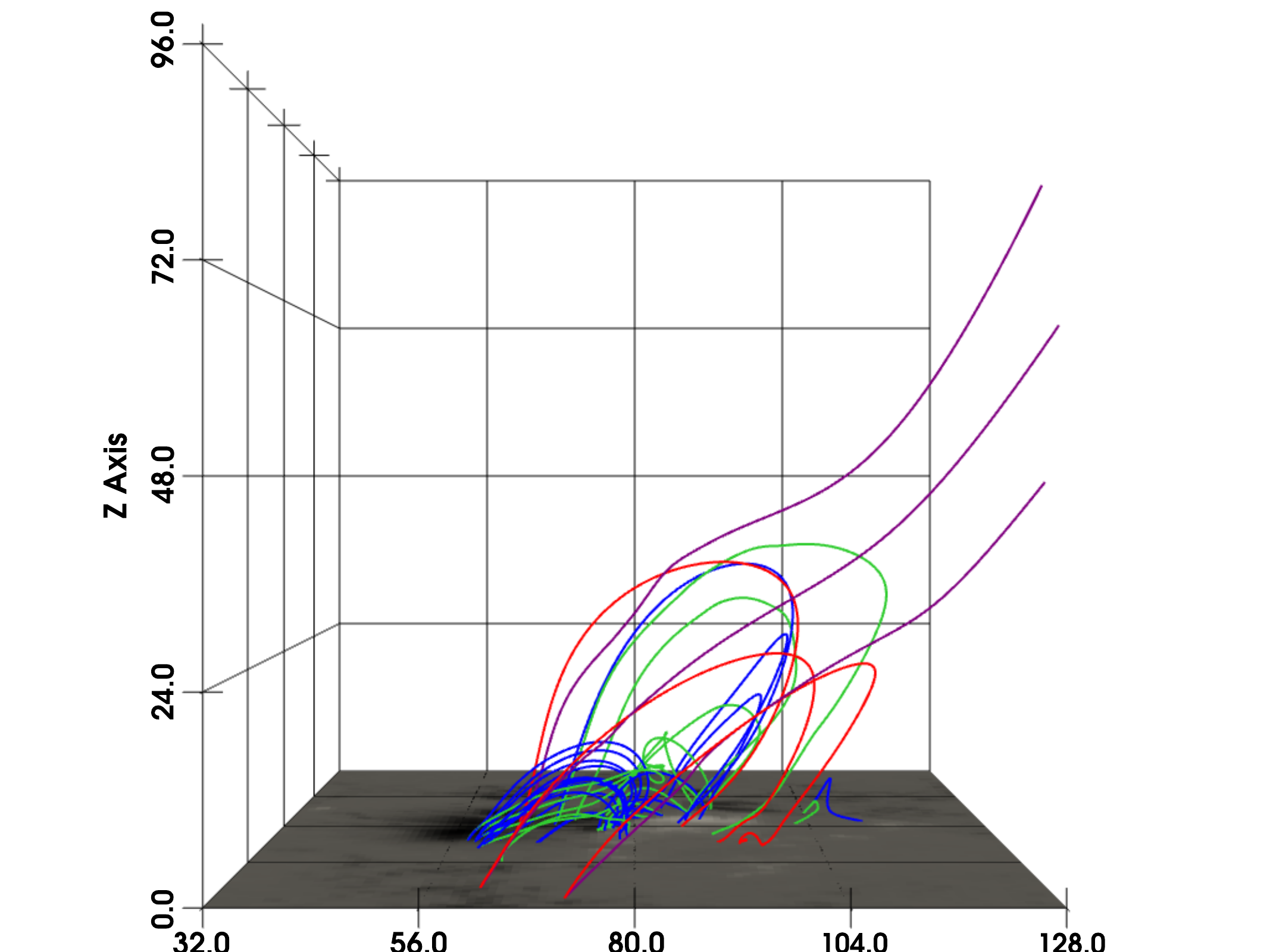}
\end{subfigure}%
\begin{subfigure}{}
    \centering
    \includegraphics[width=0.375\textwidth]{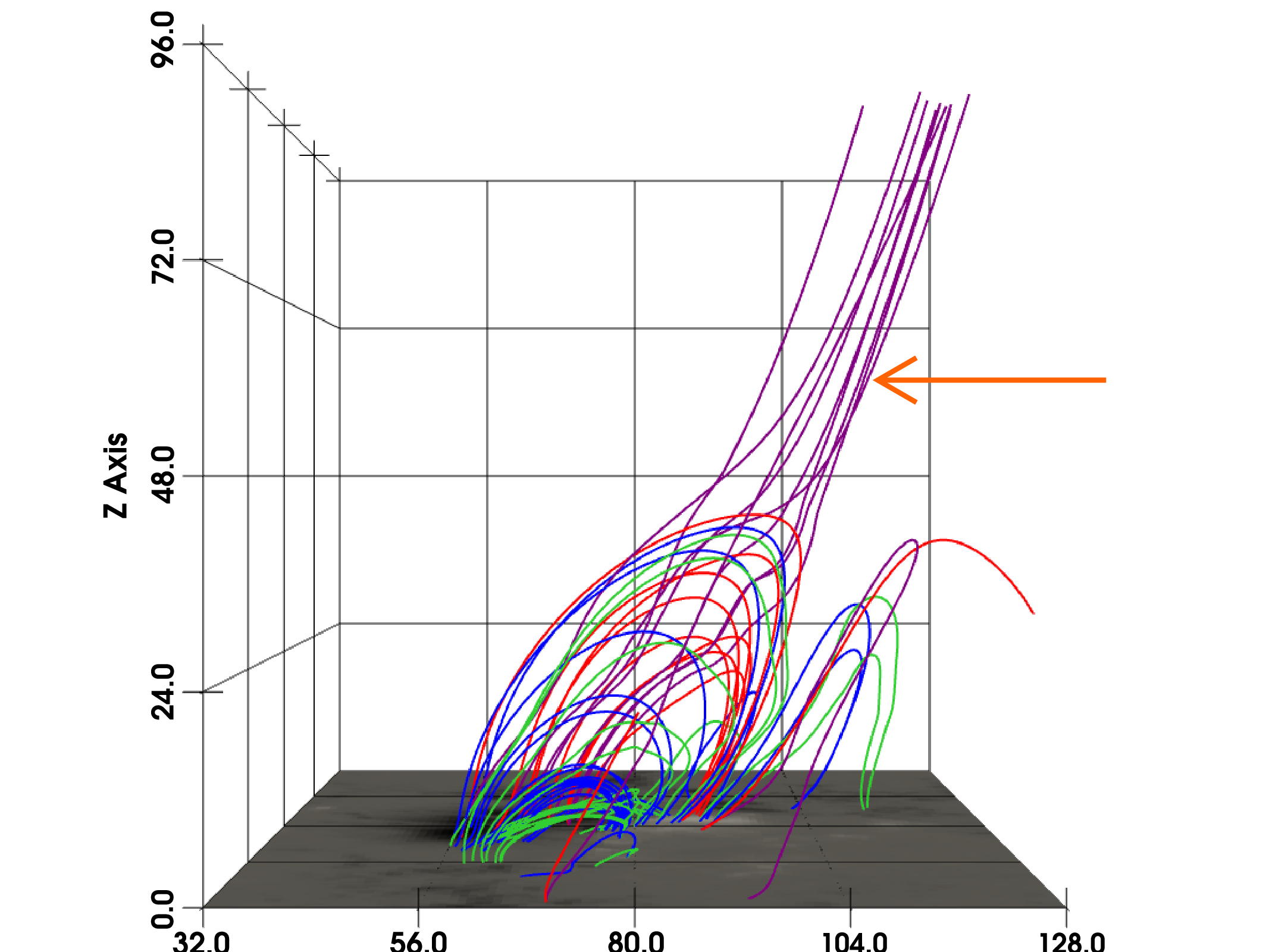}
\end{subfigure}

\begin{subfigure}{}
    \centering
    \includegraphics[width=0.05\textwidth,trim={1.5cm 1cm 6.5cm 1.25cm},clip]{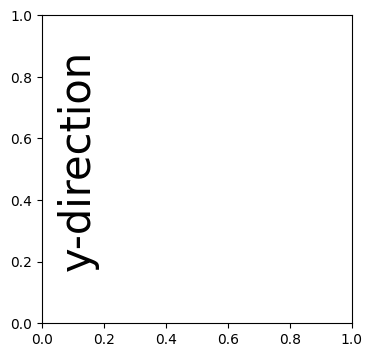}
\end{subfigure}%
\begin{subfigure}{}
    \centering
    \includegraphics[width=0.375\textwidth,trim={0cm 0.5cm 0cm 0cm},clip]{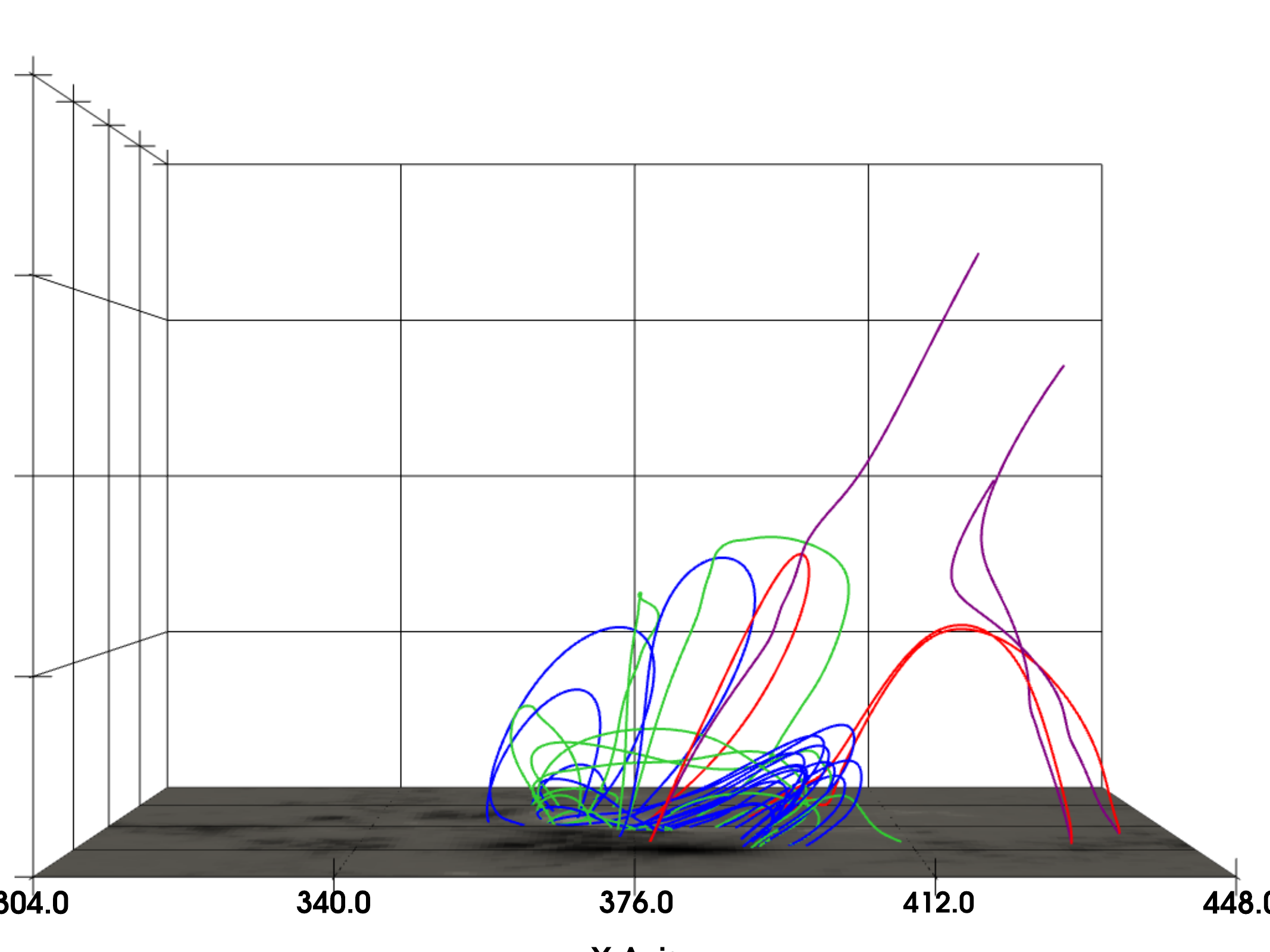}
\end{subfigure}%
\begin{subfigure}{}
    \centering
    \includegraphics[width=0.375\textwidth,trim={0cm 0.5cm 0cm 0cm},clip]{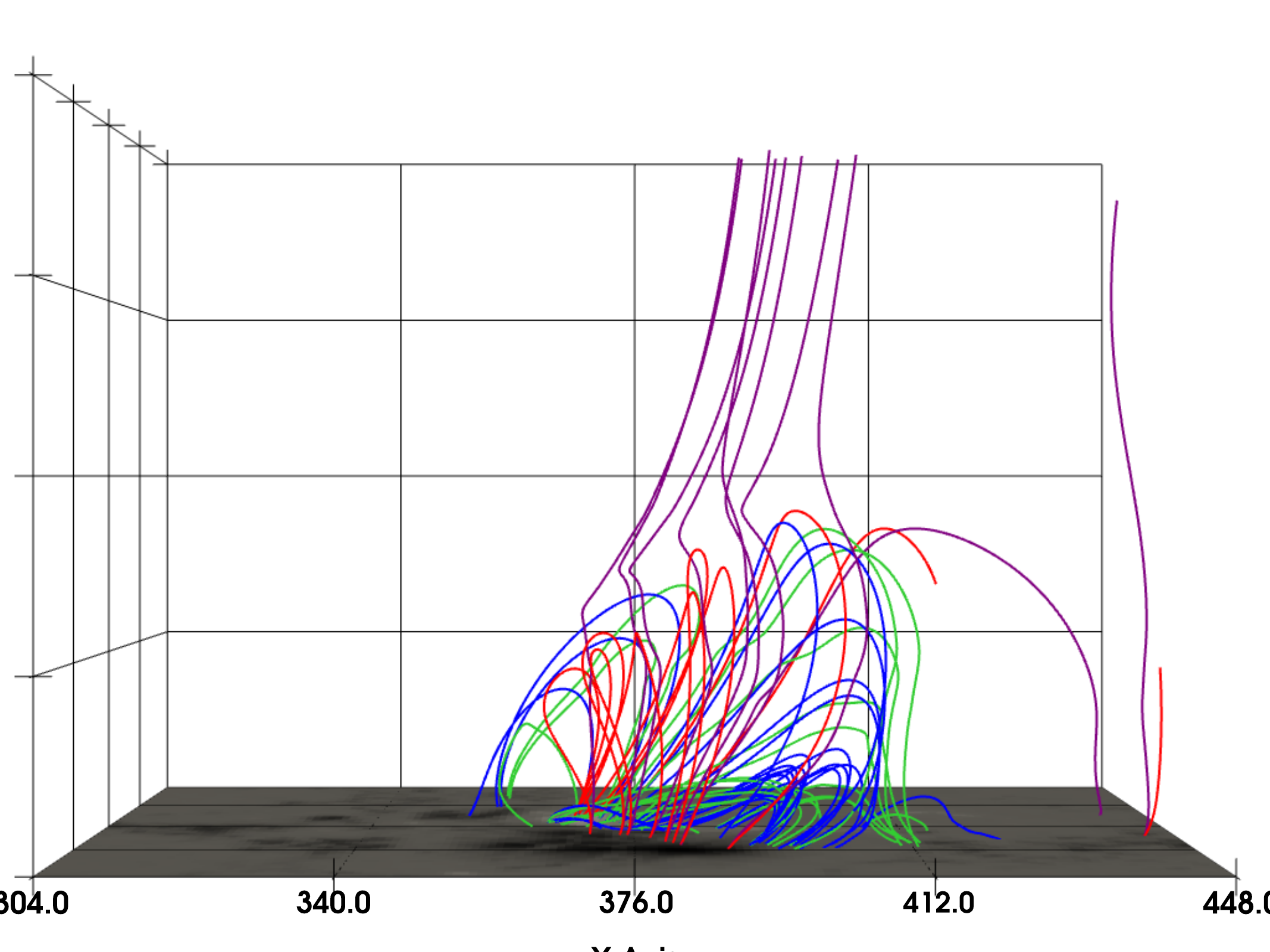}
\end{subfigure}

\begin{subfigure}{}
    \centering
    \includegraphics[width=0.05\textwidth,trim={1.5cm 1cm 6.5cm 1.25cm},clip]{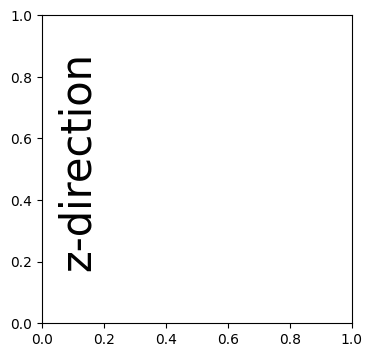}
\end{subfigure}%
\begin{subfigure}{}
    \centering
    \includegraphics[width=0.375\textwidth]{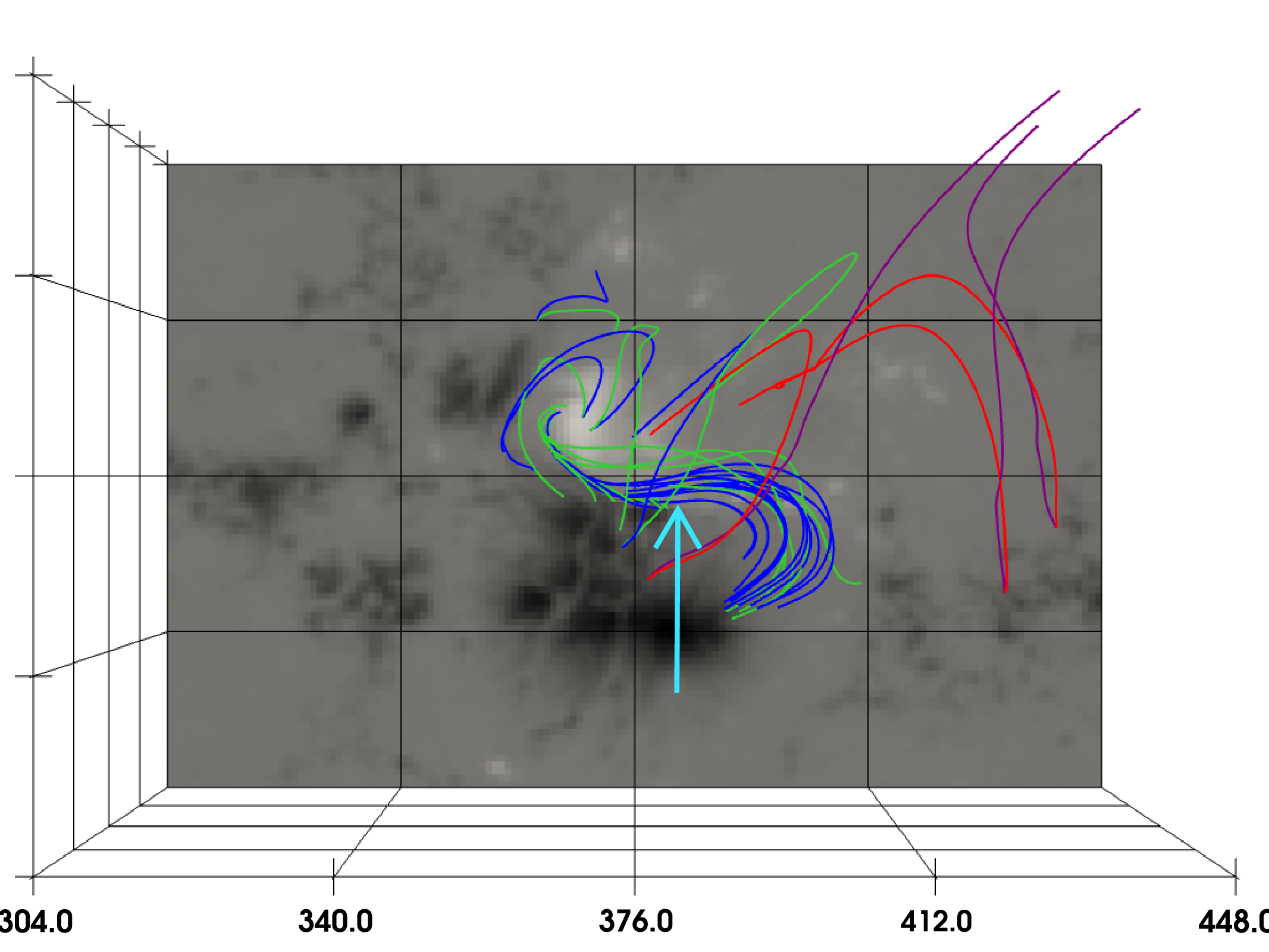}
\end{subfigure}
\begin{subfigure}{}
    \centering
    \includegraphics[width=0.375\textwidth]{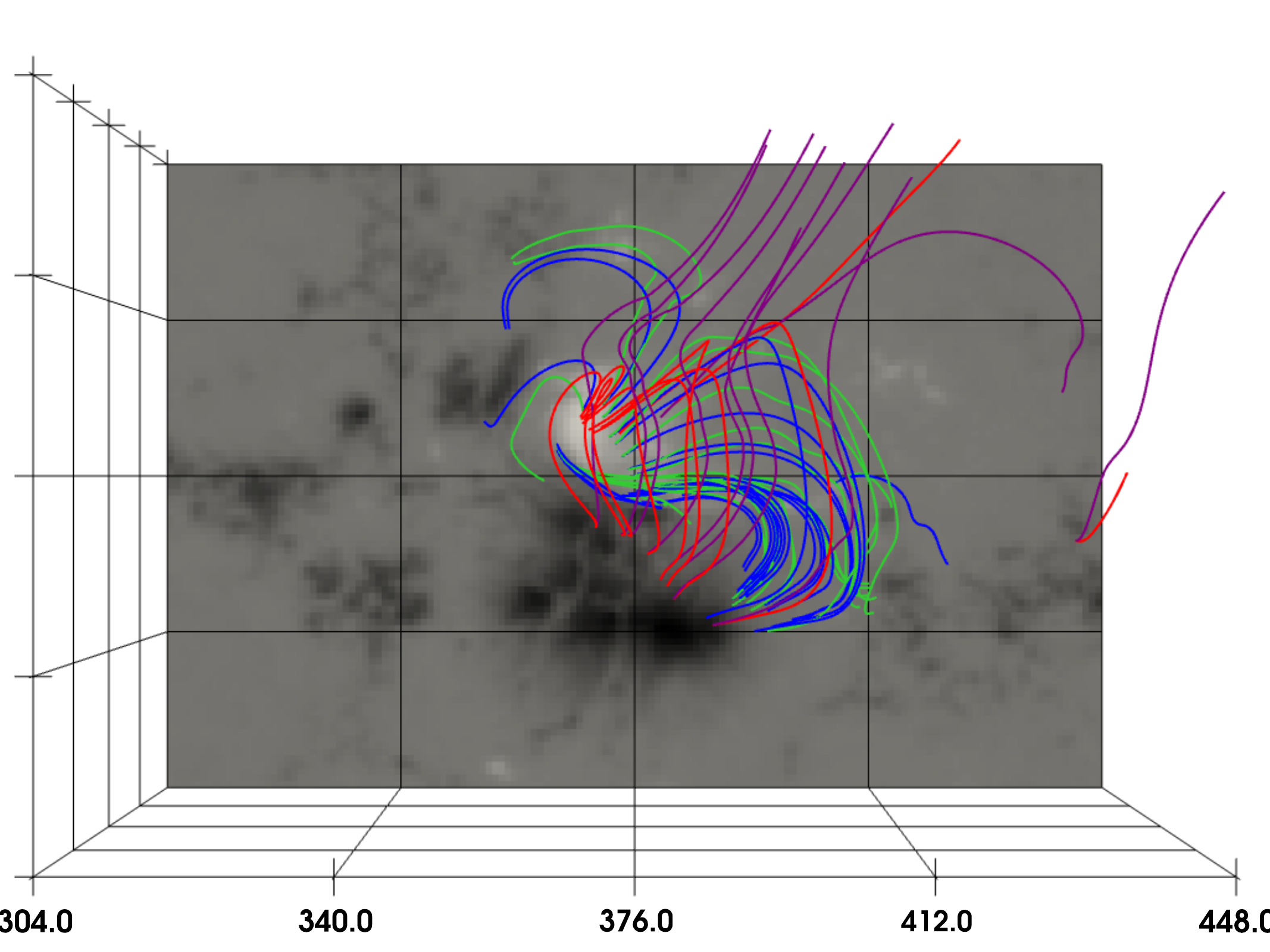}
\end{subfigure}%

\caption{Magnetic field restructuring in the two simulations. The left column shows the NLFF case, and the right column shows the non-force-free case. All panels correspond to the same times, with field line connectivity compared between $t=0$ and $t=100t_{A}$. The rows show different viewing angles: isometric, and projections along the $x$-, $y$-, and $z$-axes. Red field lines indicate structures that are initially closed at $t=0$, while the corresponding purple lines show their connectivity at $t=100t_{A}$, highlighting field lines that become open. Blue field lines denote structures at $t=0$ whose footpoints undergo significant displacement ($>5$Mm), with their later positions shown in green at $t=100t_{A}$.
}

\label{fig:restructure}
\end{figure*}

\begin{figure*}[th]
\centering
\begin{subfigure}{}
    \centering
    \includegraphics[width=0.26\textwidth]{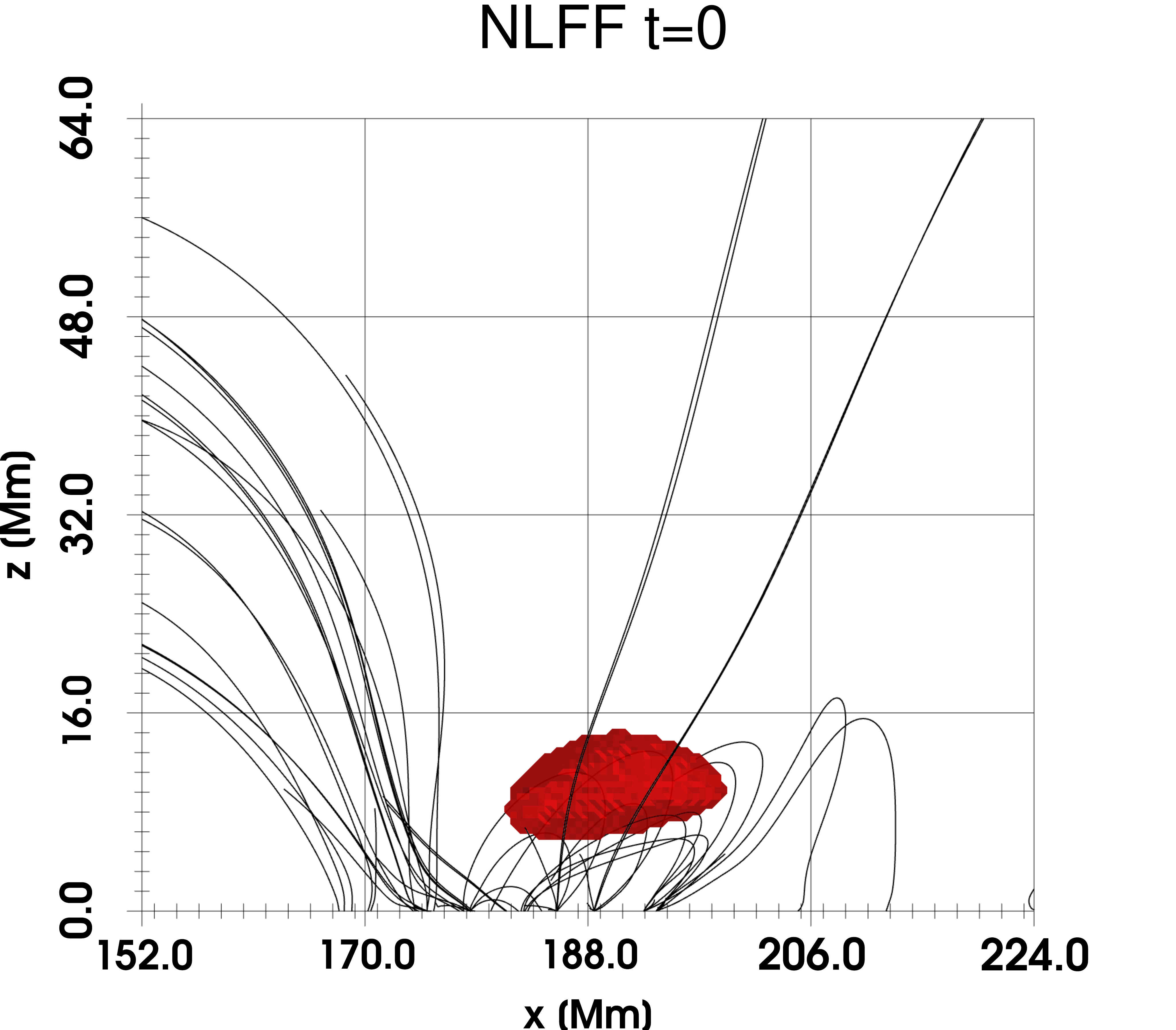}
\end{subfigure}%
\begin{subfigure}{}
    \centering
    \includegraphics[width=0.26\textwidth]{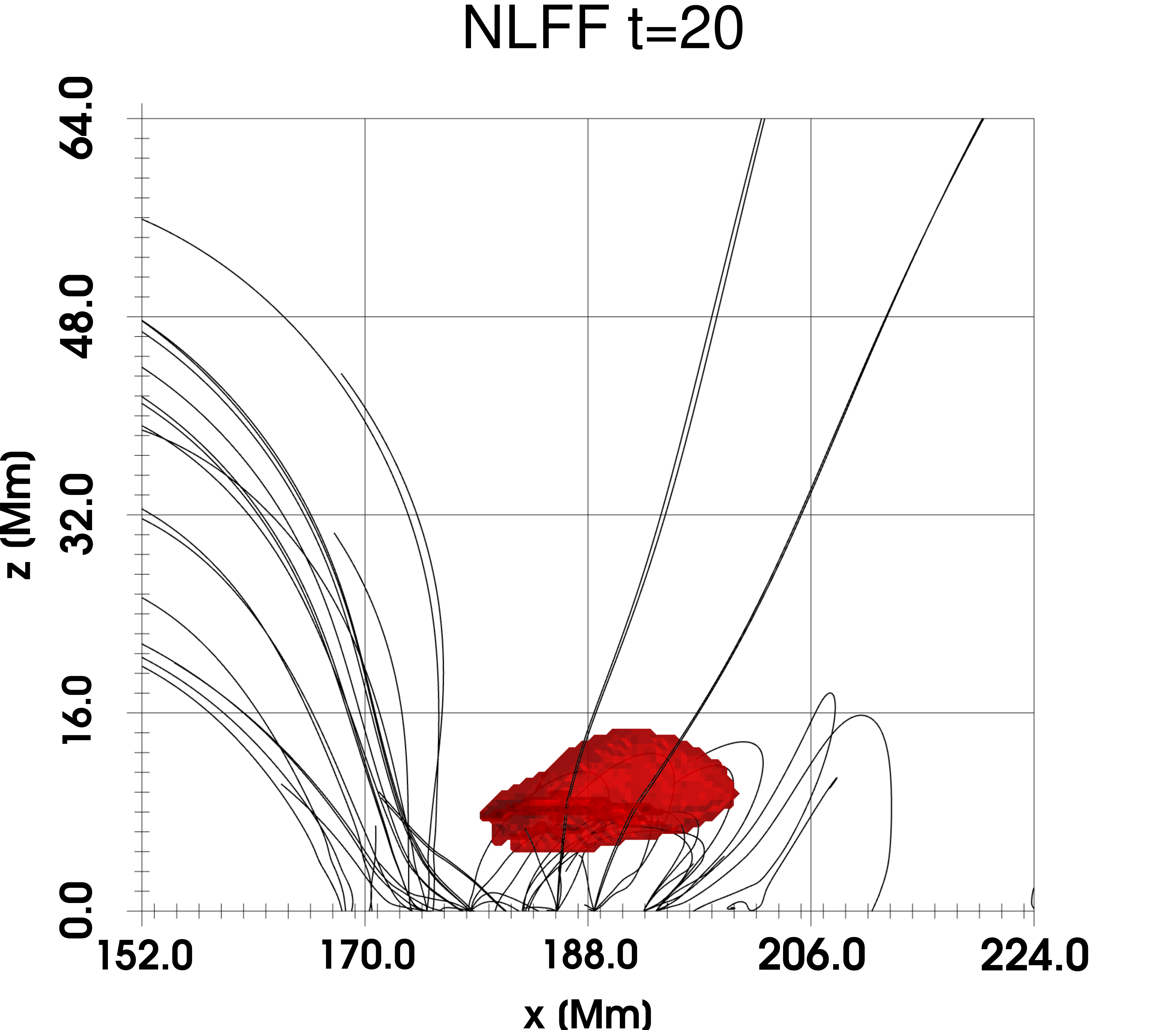}
\end{subfigure}
\begin{subfigure}{}
    \centering
    \includegraphics[width=0.26\textwidth]{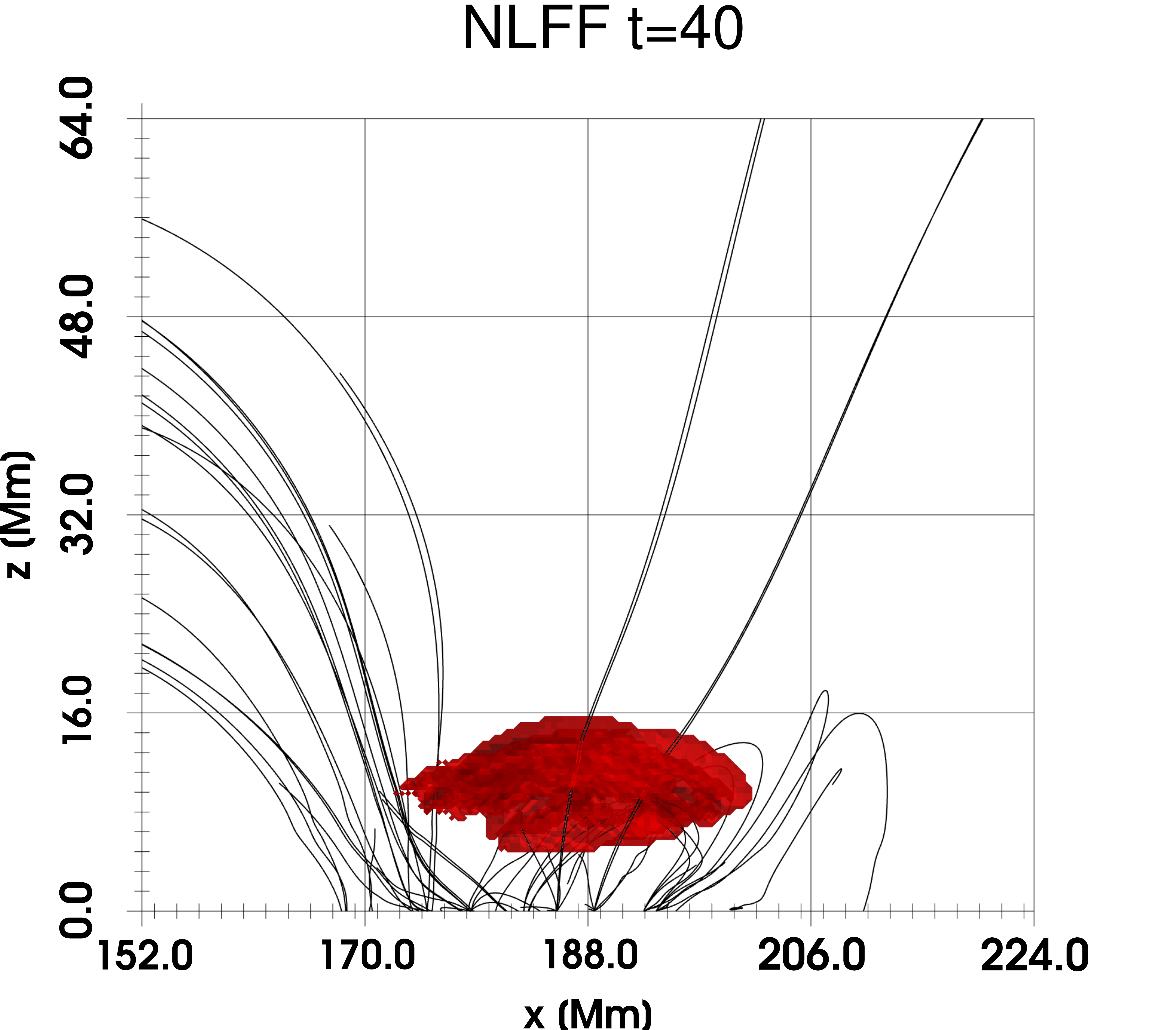}
\end{subfigure}
\begin{subfigure}{}
    \centering
    \includegraphics[width=0.26\textwidth]{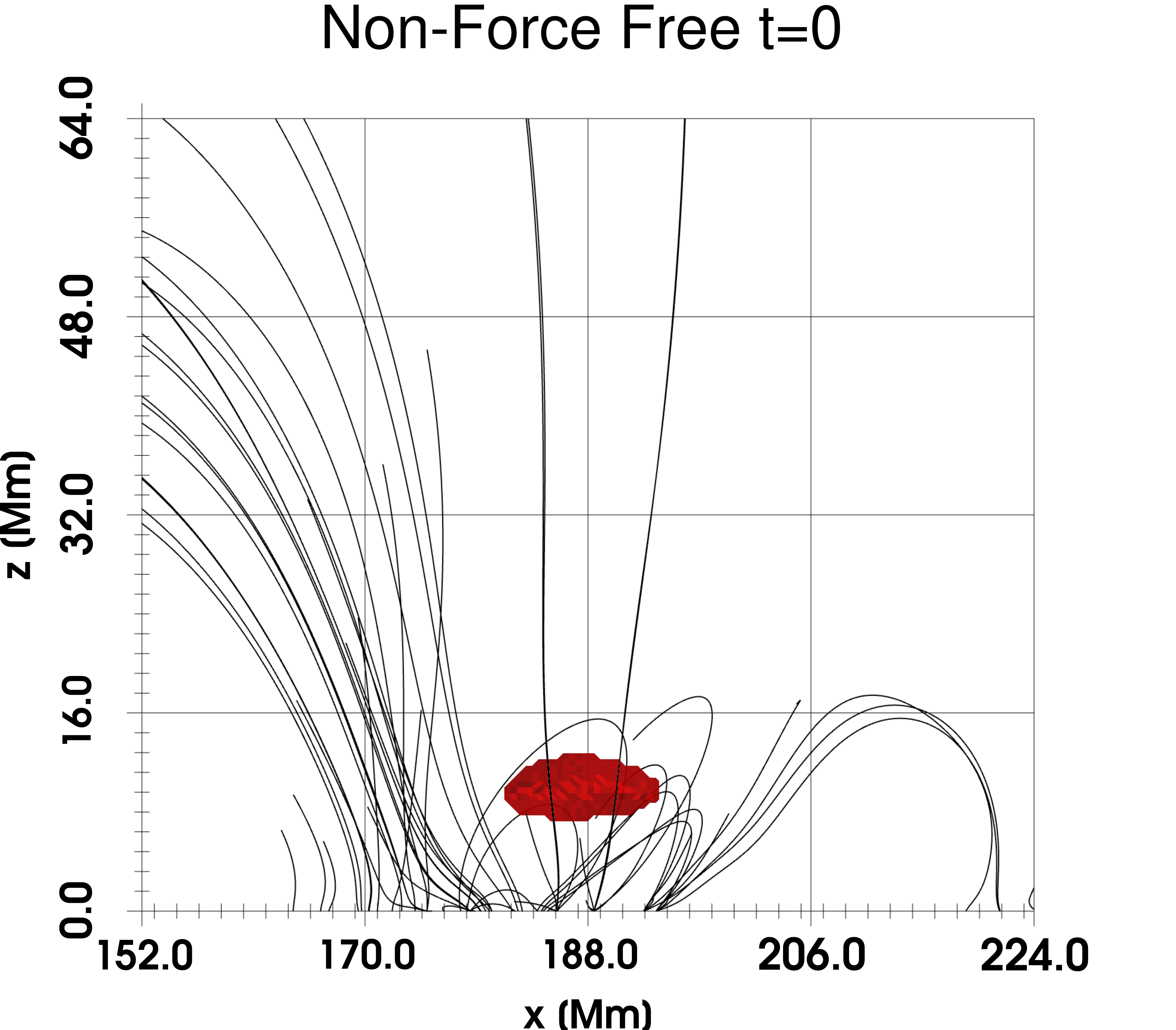}
\end{subfigure}%
\begin{subfigure}{}
    \centering
    \includegraphics[width=0.26\textwidth]{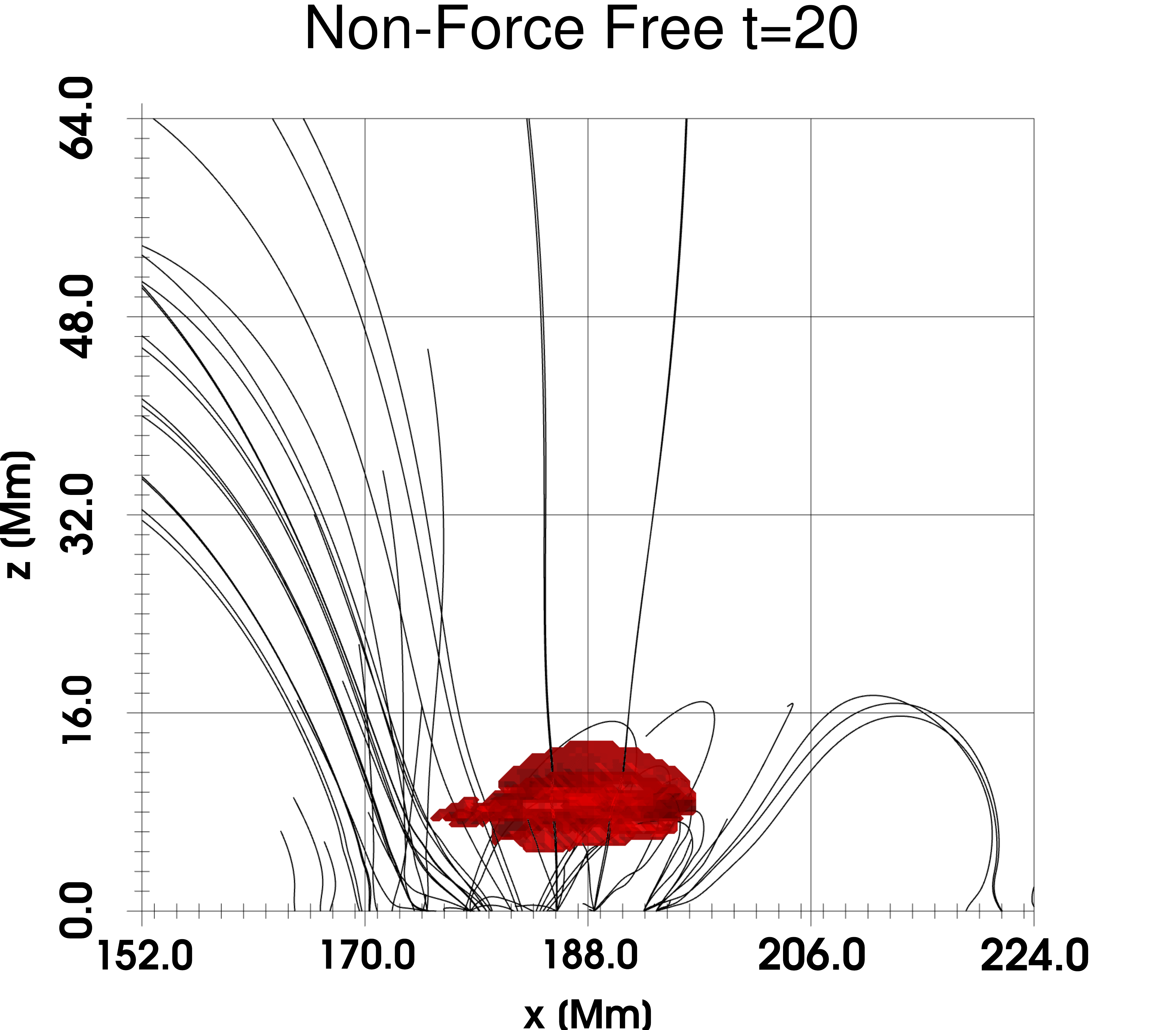}
\end{subfigure}
\begin{subfigure}{}
    \centering
    \includegraphics[width=0.26\textwidth]{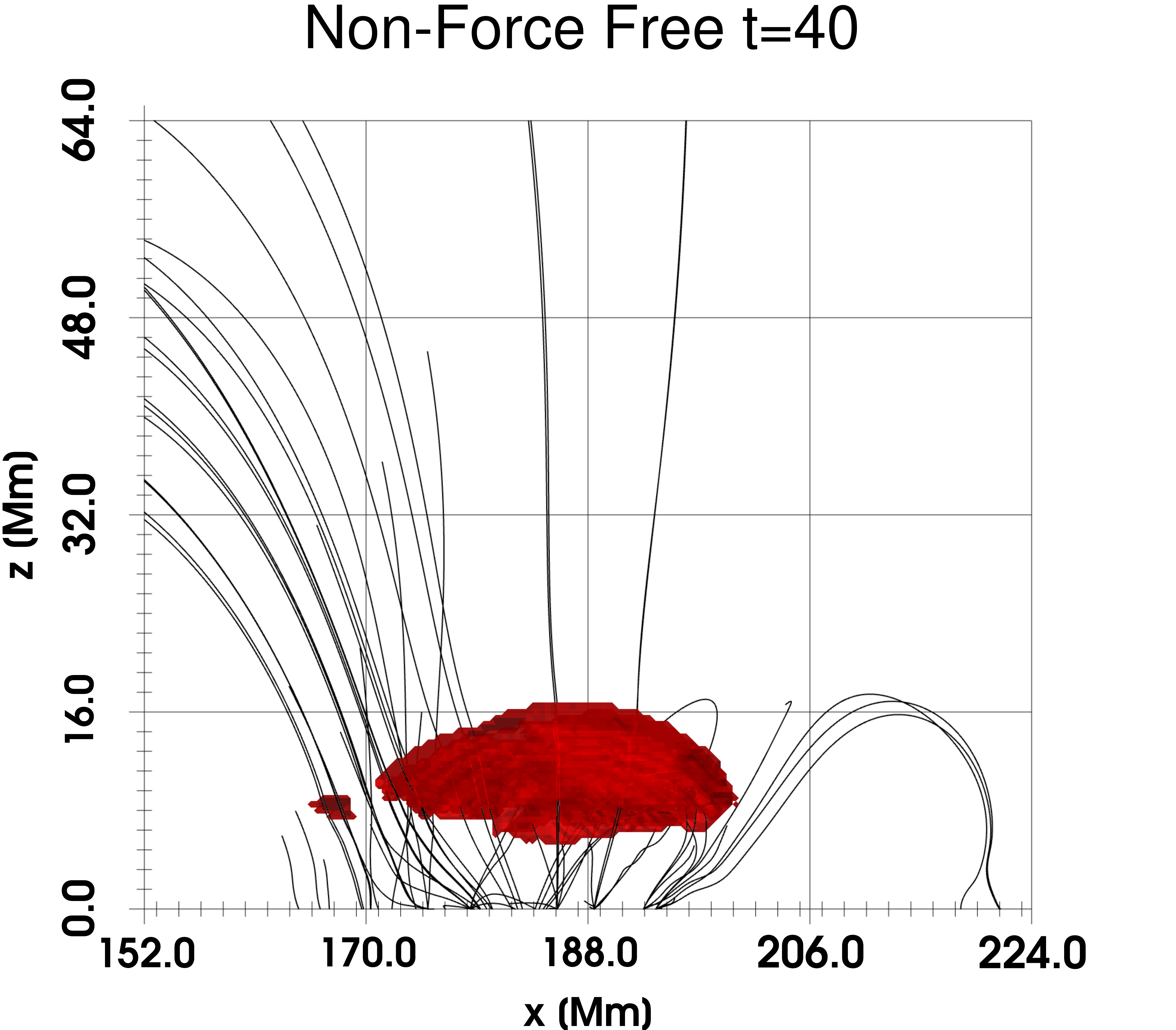}
\end{subfigure}
\begin{subfigure}{}
    \centering
    \includegraphics[width=0.36\textwidth,trim={0 0 0 1cm},clip]{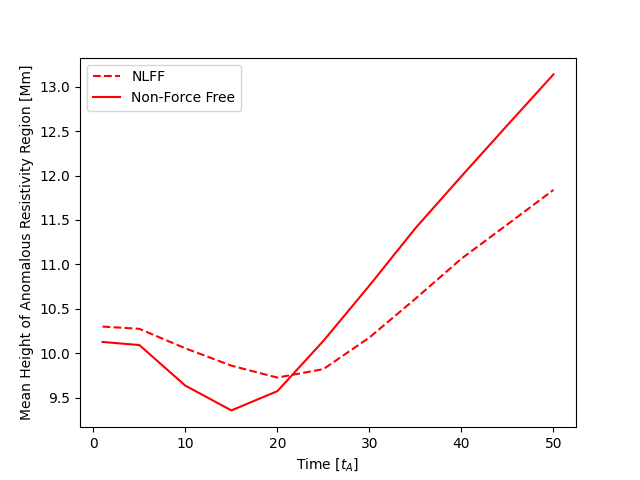}
\end{subfigure}

\caption{In the top row, results of the simulation initialised with the NLFF extrapolation is shown and in the middle row, the non-force free. The left column shows the initial configuration, the next two columns show the simulation at $t=20,40$~$t_{A}$. The magnetic field lines shown are seeded at the same locations in both, with the field line density proportional to magnetic field strength at the lower boundary. The red volume corresponds to the location where anomalous resistivity is active during that stage of the simulation. The bottom panel shows the mean height of the anomalous resistivity region in both simulations over time, with the NLFF and non-force free models being represented by the dashed and solid lines respectively.}
\label{fig:anom_bubble}

\end{figure*}




The Lorentz force summed across each layer is shown as a function of height for the NLFF and non-force free field extrapolations in the top panel of Figure~\ref{fig:lorentz}. This is presented in normalised units calculated from $L_{0},\rho_{0},B_{0}$ following \citet{arber_staggered_2001}. These are shown on a log-scale. As expected, the NLFF contains less Lorentz force than the non-force free extrapolation, and both decay with height. The NLFF extrapolation, despite the name, is not completely force free. This is due to the fact that the force is not numerically set to zero during the extrapolation. The magnetic field is reached through a magnetic relaxation process followed as described in Section~\ref{sub:NLFF}. Another factor is the lower boundary condition requiring a match with the photospheric magnetogram, which itself is not force free. In the bottom panel, this force is recast as average acceleration per layer using the stratified density profile shown in Equation~\ref{eqn:density_prof}. Neither simulation contains significant initial acceleration due to the initial magnetic field configuration, meaning that these configurations are stable enough for the study of flares.

We note that the same plane-parallel stratified atmosphere is adopted for both simulations in order to isolate the impact of the magnetic field configuration on the subsequent evolution. While this choice would be consistent with an entirely force-free field extrapolation, it does not enforce full magneto-hydrostatic balance for a non-force-free field. In practice, both extrapolations retain residual Lorentz forces due to the non-force-free nature of the photospheric boundary and the numerical construction methods. The non-force-free model can therefore be interpreted as representing a stressed pre-eruptive state. Constructing a fully consistent magnetohydrostatic equilibrium for such fields is beyond the scope of the present study.

\section{Results and Analysis}
\label{sec:res}

\begin{figure*}[th!]

\centering
\begin{subfigure}{}
    \centering
    \includegraphics[width=0.35\textwidth]{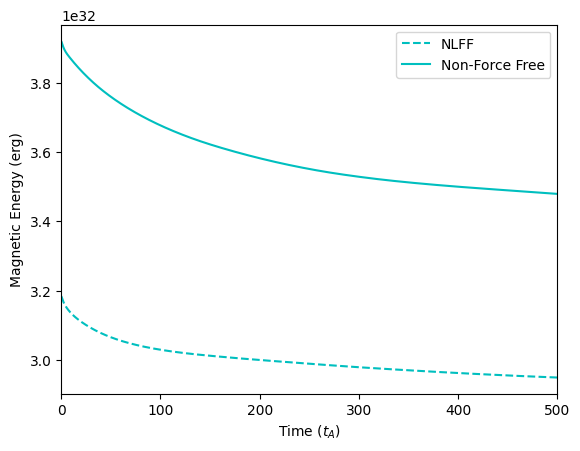}
\end{subfigure}%
\begin{subfigure}{}
    \centering
    \includegraphics[width=0.35\textwidth]{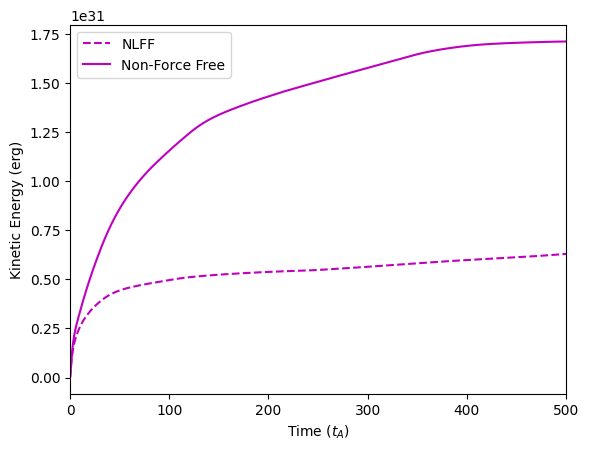}
\end{subfigure}
\begin{subfigure}{}
    \centering
    \includegraphics[width=0.35\textwidth]{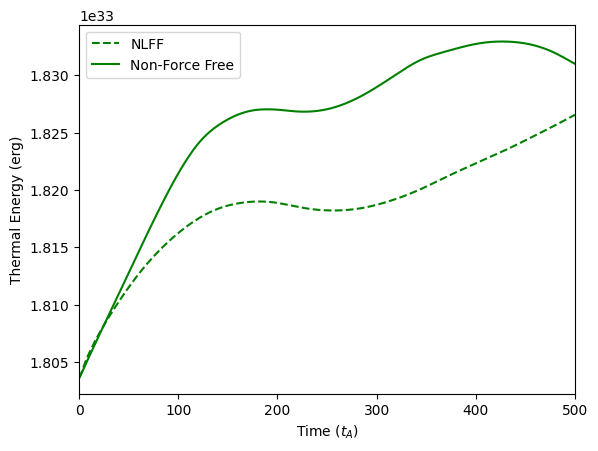}
\end{subfigure}%
\begin{subfigure}{}
    \centering
    \includegraphics[width=0.35\textwidth]{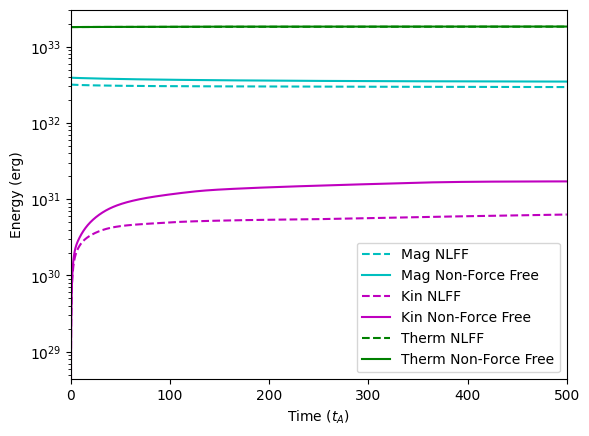}
\end{subfigure}

\caption{Energy over time for both the NLFF (dashed lines) and non-force free (solid lines) simulations. The first plot (cyan) shows magnetic energy, the second (purple) shows kinetic energy, the third (green) shows thermal energy, and the fourth shows all of them together with a log-scale for comparison.}
\label{fig:energy}
\end{figure*}

Each of these simulations is then allowed to run for $500$ nominal Alfv\'{e}n times. The magnetic field restructuring in each of these two simulations is shown in Figure~\ref{fig:restructure}. The left and right columns show the NLFF and non-force free cases respectively. The top row shows an isometric view and the next three rows show views from the $x$-, $y$-, and $z$-directions respectively. The red lines show field lines which are initially closed at $t=0$, which become open at $t=100t_{A}$ and are shown in purple. The blue lines show field lines at $t=0$ whose end points move by more than $5$~Mm across the solar surface to the green field lines shown at $t=100t_{A}$. 

Initially field lines are seeded at the same locations in both of the simulations. However, many more field lines are shown in the non-force free case as only those field lines which have significantly changed connectivity are plotted. This indicates a larger shift/restructuring of the magnetic field in the non-force free simulation. Further, there are a larger number of red $\rightarrow$ purple (closed $\rightarrow$ open) field lines in the non-force free case. This magnetic field restructuring shows a twisted sigmoid-shaped magnetic flux tube (blue $\rightarrow$ green as the flux rope untwists) which is denoted by the cyan arrow in Figure~\ref{fig:restructure}. This sigmoid structure is seen at the same location in both simulations. Also shown are an overlying set of red field lines which open into the upwards purple field lines by the end of the event, indicated by the orange arrow in Figure~\ref{fig:restructure}.

The initial volume in which the anomalous resistivity is active and its evolution is shown for the NLFF and non-force free simulations in the top and middle rows of Figure~\ref{fig:anom_bubble} respectively. This shows the progression through $t=0,20,40$~$t_{A}$ and the anomalous resistivity volume can be seen to increase in height and volume throughout time in both simulations. The average height of this anomalous resistivity volume is tracked in the bottom panel for both simulations as a function of time. This starts higher in the NLFF model and rises, but more slowly than in the non-force free model. It should be noted that a region with active anomalous resistivity  is not all that is required for reconnection to occur. A sufficiently high value of $J^{2}$ is also required. The non-force free model shows greater magnetic field line restructuring due to its higher values of $J^{2}$, despite the size and shape of the anomalous resistivity region being roughly similar in both simulations.

The evolution of different types of energy within the flare simulations are shown in Figure~\ref{fig:energy} with the NLFF shown by the dashed lines and the non-force free by the solid. The NLFF magnetic field initially contains $3.18\times10^{32}$~erg of magnetic energy, losing $2.3\times10^{31}$~erg; and the non-force free field initially contains $3.92\times10^{32}$~erg, and loses $4.4\times10^{31}$~erg. These are comparable to the expected energy releases due to an X-class flare \citep{2014ApJ...787...32M}. The NLFF field also starts with less magnetic energy, as would be expected of a force free field. The energy imparted in the form of thermal energy is of a similar order of magnitude, with an increase of $2.3\times10^{31}$~erg in the NLFF case and $2.7\times10^{31}$~erg in the non-force free case. This suggests that non-force free initialisations may help to alleviate the difficulty that NLFF initialised models have providing as much magnetic energy as is needed to produce the amount observed to be released in flares, which has been noted by previous works \citep[e.g.][]{gordovskyy_forward_2020}. Note that the energies are not exactly balanced, due to the `open' boundary condition at the top of the domain.



\begin{figure*}[th!]
    \begin{interactive}{animation}{Pub_94_3.mp4}
    \centering
    \includegraphics[width=0.75\textwidth,trim={0 0 0 0}]{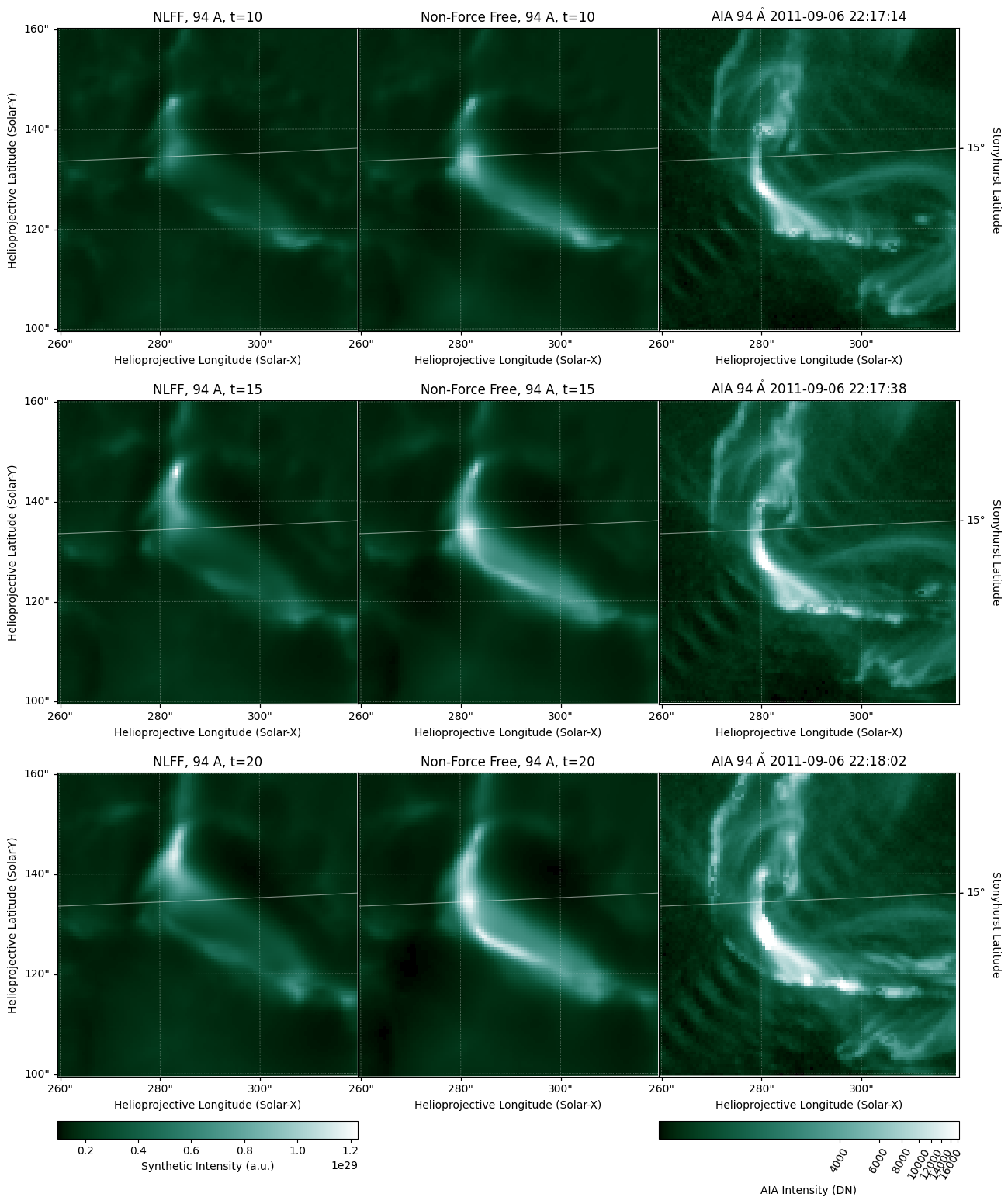}
    \end{interactive}
    \caption{A comparison of the synthetic and measured $94$~\AA~emission. In the left and central columns, synthetic emission is shown from the NLFF and non-force free initialised simulations respectively. The top row shows the simulations at $t=10t_{A}$, the middle row at $t=15t_{A}$, and the bottom row at $t=20t_{A}$. The synthetic emission is scaled to the brightest emission of the six panels (at $t=20t_{A}$ in the non-force free simulation) indicated by the lower left colour bar. The right column shows the measured AIA $94$~\AA~emission, with the top, middle, and bottom panels taken at 22:17:14, 22:17:38, and 22:18:02~UT respectively. Each measured intensity image is scaled to the brightest intensity in the latest unsaturated image, indicated by the colour bar in the lower right. An animation of this figure is available in the online version of the article. The animation contains three sections: (1) the synthetic emission from the NLFF simulation from $t=0$-$100t_{A}$; (2) the synthetic emission from the non-force free simulation from $t=0$-$100t_{A}$; (3) the measured AIA $94$~\AA~emission from 22:14:01-22:20:02~UT. }
    \label{fig:synth_grid}
\end{figure*}

Synthetic AIA $94$~\AA~emission maps produced at $t=10,15,20t_{A}$ are shown in the left and central columns of  Figure~\ref{fig:synth_grid}. These are calculated according to Section~\ref{sub:synth}. The left column shows the NLFF simulation and the central column shows the non-force free case. In the right column, the intensity measured by AIA is shown at with the same projection utilising SunPy \citep{community_sunpy_2020}. The final of these images is taken at 22:18:02~UT. This timestep is chosen as it is the final AIA image before the detector is saturated by the flare, although an animation of the AIA $94$~\AA~images is available in the online version of this article. Both simulations recreate the location of greatest emission in a similarly shaped curved bright feature. However, the brightness in the NLFF simulation is concentrated in the top-left of the structure. This is not seen in the non-force free case, where the entire structure is bright, similar to the observed case shown in the lower panel. Note that we do not expect the flare ribbons visible in the AIA observations to be visible within the simulations, due to the lack of a realistic chromosphere. 

An animation of the synthetic emission shown in the left and central columns of Figure~\ref{fig:synth_grid} is available in the online version of this article. The previously mentioned differences in structure between the two simulations are visible. Further, an eruption is visible in both simulations moving towards the top-right of the field-of-view. This eruption is also visible in the observed AIA emission. The authors suggest that this is hot material travelling along the newly open purple field lines shown in Figure~\ref{fig:restructure} by the orange arrow.

When examining the timescale of these flare simulations, it is important to note the units being considered. The Lare3d code is normalised according to three quantities and their associated normalisation constants: length ($L_{0}$), density ($\rho_{0}$), and magnetic field strength ($B_{0}$). Two of these parameters are set in relation to the original surface magnetogram ($L_0=1.0$~Mm, $B_{0}=0.1$~T). However, $\rho_{0}$ is not fixed, as the density profile of the stratified atmosphere is not measured. The normalisation constant for time is the nominal Alfv\'{e}n time, calculated in terms of the declared constants as

\begin{equation}
    t_{0}=t_{A}=\frac{L_{0}\sqrt{\rho_{0}\mu_{0}}}{B_{0}} \ ,
\end{equation}

where $\mu_{0}$ is the magnetic permeability of free space. By considering a range of physically reasonable particle densities at the top of our $128$~Mm tall computational domain $\rho_{t}=[2\times10^{8},10^{10}]$~cm$^{-3}$, we arrive at a range of values for $t_{0}=[0.09,0.29]$~s. This range of density scalings is considered in order to best match the speed of the simulated flare to the observed flare. As this is simply a change of scaling parameters, this has no effect on the physical evolution of the simulation, only the conversion of simulation units to physical units.

\begin{figure}[t]

\centering
\includegraphics[width=0.45\textwidth]{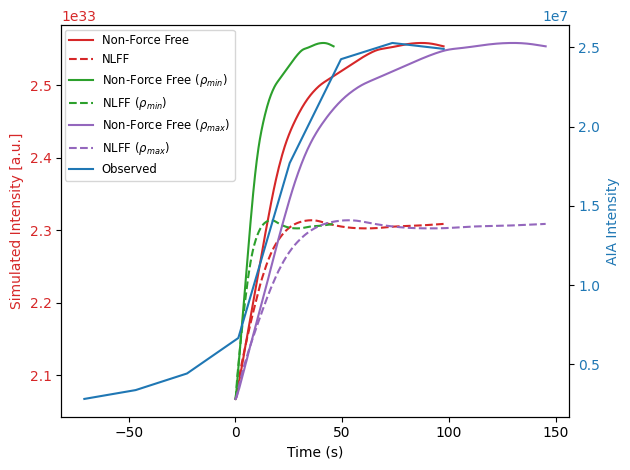}

\caption{The green, red, and purple lines show the synthetic $94$~\AA~light curves for both of the simulations with density scaling factors corresponding to desity at the upper domain boundary of $\rho_{t}=2\times10^{8},9\times10^{8},$ and $10^{10}$~cm$^{-3}$ respectively. The solid lines corresponds to the non-force free case, and the dashed lines to the NLFF case. The blue line shows the measured AIA $94$~\AA~light curve. }
\label{fig:light_curve}
\end{figure}

Synthetic $94$~\AA~light curves for each of the simulations are shown in Figure~\ref{fig:light_curve} with the solid line corresponding to the non-force free case, and the dashed line to the NLFF simulation. A selection of three time scalings are shown, with the minimum and maximum values of density at the upper boundary ($\rho_{t}=2\times10^{8},10^{10}$~cm$^{-3}$) shown in green and purple respectively. Also shown for comparison is the $94$~\AA~light curve measured by AIA. Finally, a `tuned' density scaling parameter is selected to match the timescale of the observed AIA light curve, shown in red and corresponding to a value of $\rho_{t}=9\times10^{8}$~cm$^{-3}$.  Although both simulations show a similar temporal evolution, the non-force free case is brighter, as expected due to the larger energy release in this simulation. This discrepancy in brightness is similarly visible in Figure~\ref{fig:synth_grid}.

\section{Conclusions} 
\label{sec:conc}

In this work, we have performed a controlled comparison of data-driven MHD flare simulations initialised with either a conventional NLFF magnetic field extrapolation or a non-force free magnetic field model that self-consistently incorporates plasma forces. Although the two extrapolation techniques produce broadly similar large-scale magnetic structures prior to flare onset, we find that their subsequent magnetic reconnection, energy release, and flaring evolution differ substantially. In particular, the non-force free model undergoes more extensive magnetic restructuring and releases significantly more magnetic energy during the flare, alleviating the long-standing difficulty of NLFF-initialised simulations underestimating the energy budget of observed X-class flares.

These differences in magnetic evolution have clear observational consequences. Synthetic EUV emission calculated for the AIA $94$~\AA~channel reveals marked contrasts between the two simulations, with the non-force free model producing a brightness distribution and flaring loop structure that more closely resemble the observed flare emission. In contrast, the NLFF-initialised simulation exhibits a more localised and less realistic emission pattern. This demonstrates that assumptions made in constructing the initial coronal magnetic field can propagate directly into observable flare signatures.

Finally, our results highlight the diagnostic power of synthetic EUV emission in assessing the realism of data-driven flare simulations. Direct comparison between simulated and observed emission provides a stringent and physically meaningful test of competing magnetic field models. Taken together, these findings suggest that non-force free extrapolations offer a promising pathway toward more realistic flare simulations and motivate their broader adoption in future data-constrained MHD studies of solar eruptive events.

\section*{Data Access Statement}

Due to the data volume (approaching 3 TB), the MHD simulation data is available, in various stages of reduction format, directly from the first author (William Bate; w.bate@herts.ac.uk).

\vspace{20mm}

\section*{Acknowledgements}
W.B. and M.G. were supported by the Science and Technology Facilities Council [grant number ST/Y001141/1]. W.B. is appreciative of the DiRAC Seedcorn Project dp416. A.P. acknowledges the support from the Research Council of Norway through its Centres of Excellence scheme, project number 262622. A.P., A.S.B, A.S, and M.V.S. acknowledge the Synergy Grant number 810218 (“The Whole Sun” ERC-2018-SyG) of the European Research Council. A.S., A.S.B. and M.V.S. acknowledge the French Agence Nationale de la Recherche (ANR) project STORMGENESIS \#ANR-22-CE31-0013-01, the CNRS-INSU AT-ST and CNES Solar Orbiter grant for financial support.

This work used the DiRAC Memory Intensive service (Cosma8) at Durham University, managed by the Institute for Computational Cosmology on behalf of the STFC DiRAC HPC Facility (www.dirac.ac.uk). The DiRAC service at Durham was funded by BEIS, UKRI and STFC capital funding, Durham University and STFC operations grants. DiRAC is part of the UKRI Digital Research Infrastructure. 

CHIANTI is a collaborative project involving George Mason University, the University of Michigan (USA), University of Cambridge (UK) and NASA Goddard Space Flight Center (USA). 

%

\vspace{5mm}
\facilities{HMI \citep[][]{2012SoPh..275..327S}, AIA \citep[][]{2012SoPh..275...17L}, DiRAC Cosma8}





\bibliography{apjbib}{}
\bibliographystyle{aasjournal}

\end{document}